\documentclass[a4paper,twocolumn,aps,prd,eqsecnum,nofootinbib,balancelastpage,notitlepage]{revtex4-2}

\usepackage[dvipdfmx,hiresbb]{graphicx}% Include figure files
\usepackage{bm}% bold math
\usepackage{amsmath}
\usepackage{amssymb}
\usepackage{amsfonts}
\usepackage{color}
\newcommand{\half}{\frac{1}{2}}

\newcommand{\der}{\partial}

\newcommand{\tr}{\mbox{\rm tr}}

\newcommand{\dsl}{\partial\kern-0.55em\raise 0.14ex\hbox{/}}
\newcommand{\pd}[2]{\frac{\partial #1}{\partial #2}}
\newcommand{\xT}{x_{\text{T}}}
\newcommand{\xF}{x_{\text{F}}}
\newcommand{\xcl}{x_{\text{cl}}}
\newcommand{\xR}{x_{\text{R}}}
\newcommand{\xb}{x_{\text{b}}}
\newcommand{\xr}{x_{\text{r}}}
\newcommand{\xB}{x_{\text{B}}}
\newcommand{\dotxB}{\dot{x}_{\text{B}}}
\newcommand{\ddotxB}{\ddot{x}_{\text{B}}}
\newcommand{\SB}{S_{\text{B}}}
\newcommand{\zb}{z_{\text{B}}}
\newcommand{\xsB}{x_{\text{sB}}}
\newcommand{\dotxsB}{\dot{x}_{\text{sB}}}
\newcommand{\ddotxsB}{\ddot{x}_{\text{sB}}}

\newcommand{\ZB}{Z_{\text{B}}}
\newcommand{\SsB}{S_{\text{sB}}}
\newcommand{\xsF}{x_{\text{sF}}}
\newcommand{\dotxsF}{\dot{x}_{\text{sF}}}
\newcommand{\ddotxsF}{\ddot{x}_{\text{sF}}}
\newcommand{\SsF}{S_{\text{sF}}}
\newcommand{\ZsF}{Z_{\text{sF}}}
\newcommand{\xs}{x_{\text{s}}}
\newcommand{\xt}{x_{\text{t}}}
\newcommand{\xC}{x_{\text{C}}}
\begin{document}

%Title of paper
\title{Saddle-point approximation to the false vacuum decay at finite temperature in one-dimensional quantum mechanics
}

\author{Koji Harada}
\email{harada@artsci.kyushu-u.ac.jp}
\affiliation{Faculty of Arts and Science, Kyushu University 744 Motooka, Nishi-ku, Fukuoka 819-0395 Japan}
\author{Shuichiro Tao}
\email{tao@okayama-u.ac.jp}
\affiliation{Institute of Promotion of Education and Campus Life,  Okayama University 2-1-1 Tsushima-naka, Kita-ku, Okayama 700-8530 Japan}
\author{Qiang Yin}
\email{corresponding author \\ in.kyo@phys.kyushu-u.ac.jp}
\affiliation{Graduate School of Science, Kyushu University 744 Motooka, Nishi-ku, Fukuoka 819-0395 Japan}

\date{\today}

\begin{abstract}
 We calculate the false-vacuum decay rate in one-dimensional quantum
 mechanics on the basis of the saddle-point approximation in the
 Euclidean path integral at finite temperature. The saddle points are
 the finite-$T$ and shifted bounce solutions, which are finite-period
 analogs of the (zero-temperature) bounce solution, and the shot
 solutions. We re-examined the zero-temperature result by Callan and
 Coleman and compare with the zero-temperature limit of our results. We
 also perform some numerical calculations to illustrate the temperature
 dependence of the decay rate and compare it with the result by Affleck.
\end{abstract}

\maketitle

\section{Introduction}
\label{sec:intro}

% importance of first-order phase transition
%・Electroweak baryogenesis
%・gravitational waves
%・Dark matter 
Even though the electroweak phase transition in the standard model seems
to be crossover~\cite{PhysRevLett.77.2887,DOnofrio:2015gop}, some
possible extensions of the Higgs sector may cause a first-order phase
transition and would leads to very interesting consequences. A
well-known example is the electroweak baryogenesis~\cite{Kuzmin:1985mm}
(\cite{Cohen:1993nk, Rubakov:1996vz, Funakubo:1996dw, Morrissey:2012db}
for reviews).  Recently, much attention has been paid to the
gravitational waves generated by the collision of the bubbles of the
true vacuum~\cite{PhysRevD.45.4514, PhysRevD.49.2837, Huber:2008hg}
(\cite{Weir:2017wfa} for a review) and the dark matter production due to
the filtering effect~\cite{Baker:2019ndr}, or to the ultrarelativistic
bubble expansion~\cite{Azatov:2021ifm}.

% False vacuum decay
% Coleman 
% bounce solution
If the phase transition is of the first order, the unstable ``false
vacuum'' decays into the stable ``true vacuum'' through
quantum-mechanical tunneling.  The decay rate through such tunneling
has been given by Coleman~\cite{PhysRevD.15.2929, PhysRevD.16.1248},
by using saddle-point approximation in the Euclidean path-integral
formalism. See Refs.~\cite{Kobzarev:1974cp,Langer:1967ax} for earlier
investigations.

% 1-dim QM
A one-dimensional quantum-mechanical system provides a simple example of
such a phase transition and the decay rate has been calculated by Callan
and Coleman~\cite{PhysRevD.16.1762}. The classical solution called the
``bounce solution'' plays the central role in the calculation, and the
so-called potential deformation method is employed to get the imaginary
part of the (ground-state) energy. See also Ref.~\cite{Coleman_1985}
(especially the appendices) for useful explanations, and the first few
sections of Ref.~\cite{PhysRevD.95.085011} for an excellent review.

% finite temperature
% Affleck, Linde
In many applications (in particular in cosmological applications), one
would be interested in extending the Callan-Coleman calculation to the
finite-temperature cases. Affleck~\cite{PhysRevLett.46.388} provides a
finite-temperature version of the decay rate
formula. Linde~\cite{Linde:1980tt,Linde:1981zj} also provides the
formula for very high temperature, where the field theory effectively
becomes three-dimensional. In those papers, the formulae are given as
natural extensions of the Coleman's formula, without explicitly
mentioning the classical solutions (saddle-points) at finite
temperature.

% In this paper:
In this paper, we derive the false vacuum decay rate of a
one-dimensional quantum mechanical system (particle) at finite
temperature in the saddle-point approximation by explicitly using the
(Euclidean) classical solutions. There are several subtleties which, to
our best knowledge, have not been noticed before.

% the outline is given in intro 
In the following, we outline how to calculate the vacuum decay rate. The
detailed calculations are given in the subsequent sections. It is to
introduce some basic concepts which will be important later in this
paper.

\begin{figure}[h]
 \includegraphics[width=\linewidth,clip]{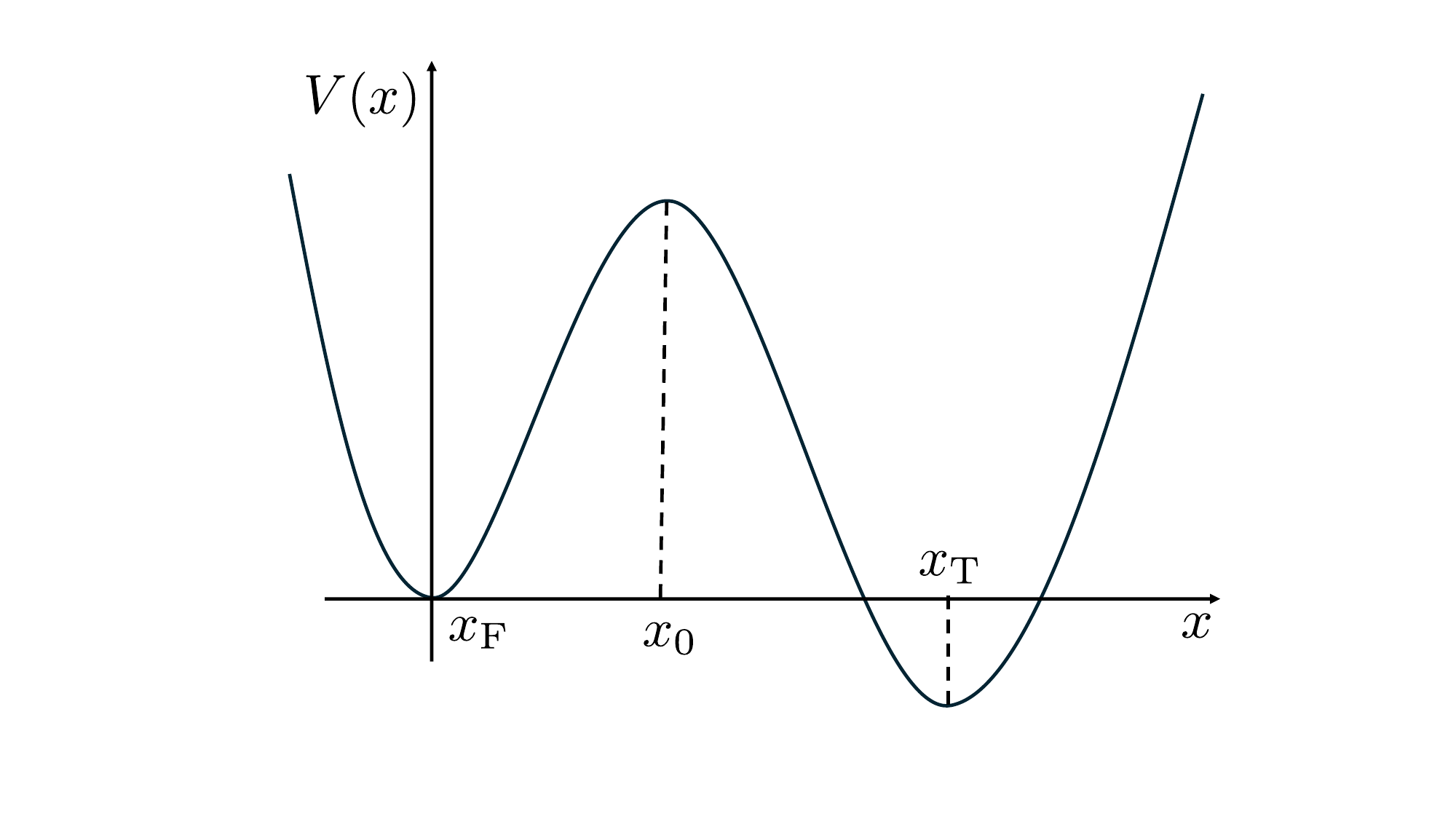}
 \caption{\label{fig:potential} A typical potential
 under consideration.}
\end{figure}

% First of all
We put a one-dimensional quantum-mechanical system, described by the
Hamiltonian
\begin{equation}
 H =\frac{1}{2m} p^2 +V(x),
\end{equation}
in contact with a heat bath, characterized by the temperature $T$. The
potential $V(x)$ (see Fig.~\ref{fig:potential} ) has two minima at
$x=x_{\text{F}}$ (false vacuum) and at $x=\xT$ (true vacuum), with
$V(\xF)=0 > V(\xT)$. The system is initially in the false-vacuum well,
i.e., around $x=\xF$, and decays into the true vacuum. We calculate this
decay rate in the equilibrium statistical mechanics framework. It means
that we assume that the system is almost in the thermal equilibrium and
the false-vacuum decay rate is very small, so that the relaxation time
is much shorter than the inverse of the decay rate.

% potential for x>x_F is irrelevant
We will see in later sections that the results are independent of the
potential shape for $x>\xT$. One may imagine that $V(x) = V(\xT)$ for
$x\ge \xT$, so that one does not need to worry about the reflecting
waves.

% what happens at finite temperatures
The basic ingredient of the finite-temperature calculation is the
partition function $Z$, or the Helmholtz's free energy $F$,
\begin{equation}
 Z= e^{-\beta F} =\tr \left(e^{-\beta H}\right),
\end{equation}
where $\beta =1/(k_{\text{B}}T)$. 

% trace and path integral
The trace may be taken in the coordinate space, so that
\begin{align}
 Z &= \int dy \;\langle y | e^{-\beta H} | y\rangle
 \notag \\
 &= \int dy\; \mathcal{N}
 \int_{x(\pm\hbar\beta/2)=y} \mathcal{D}x(\tau)\; 
 e^{-S[x]/\hbar},
 \label{partition_function}
\end{align}
where the path integral is over all the paths satisfying the boundary
condition $x(\pm \beta\hbar/2)=y$ and 
\begin{equation}
 S[x] =\int_{-\hbar\beta/2}^{\hbar\beta/2} d\tau 
  \left[
   \frac{m}{2} \left(\frac{dx}{d\tau}\right)^2+V(x)
  \right]
\end{equation}
is the Euclidean action, and $\mathcal{N}$ is a normalization constant.

% saddle points = classical solutions
We consider the saddle-point approximation to the free energy. A saddle
point is the (Euclidean) classical solution $\xcl(\tau)$,
\begin{equation}
 m\frac{d^2}{d\tau^2}\xcl(\tau) -V'(\xcl(\tau))=0,
  \label{EOM}
\end{equation}
subject to the boundary conditions;
\begin{equation}
 \xcl(\pm \beta\hbar/2) = y.
  \label{bc_cl}
\end{equation}
We expand the path $x(\tau)$ around the classical solution
$\xcl(\tau)$ and write
\begin{equation}
 x(\tau) = \xcl(\tau) +r(\tau),
  \label{def_r}
\end{equation}
where the fluctuation around the classical solution $r(\tau)$ satisfies
the Dirichlet boundary conditions,
\begin{equation}
 r(\pm \hbar \beta/2) =0.
  \label{r_bc}
\end{equation}
The saddle-point approximation consists of the exponential factor of the
classical action, and the prefactor obtained by the Gaussian integration
of the fluctuations.

\begin{figure}[h]
 \includegraphics[width=\linewidth,clip]{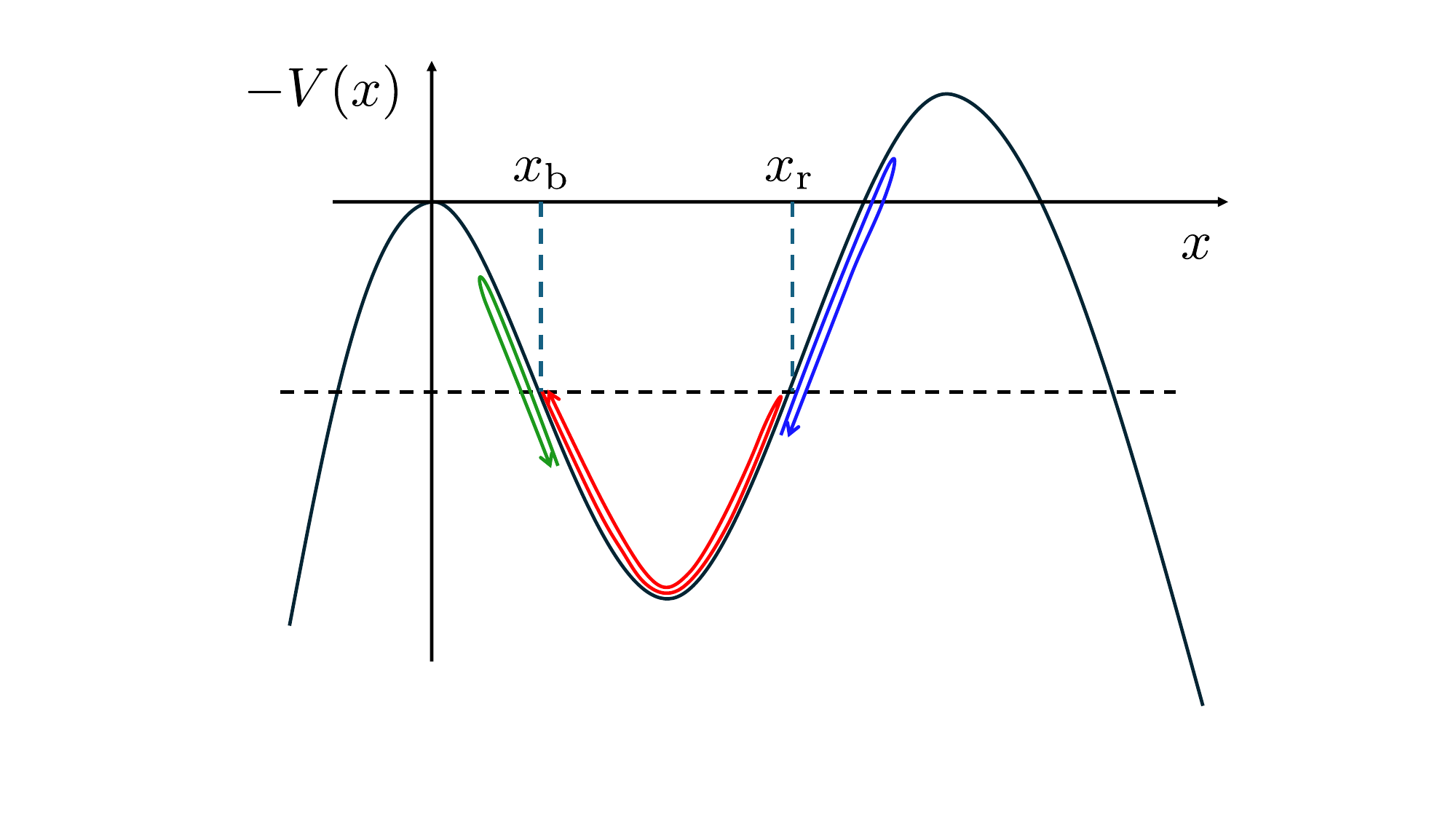}
 \caption{\label{fig:three_solutions}
 Three kinds of classical solutions. The red trajectory denotes
 the finite-$T$ bounce solution. The green and the blue trajectories
 denote shot-F and shot-T solutions, respectively. }
\end{figure}

% two classes of classical solutions
There are two classes of classical solutions which satisfy the boundary
conditions~\eqref{bc_cl}. A solution of the first class is ``periodic''
in the sense that it comes back to the original point with the same
velocity after the (Euclidean) time period $\beta\hbar$ and can repeat
the same motion again and again. On the other hand, a solution of the
second class just returns back to the original position $y$ after
$\beta\hbar$, but the velocity is not the same as before (it changes the
sign).

% DHN 
In the classical paper of Dashen, Hasslacher, and
Neveu~\cite{Dashen:1974ci}, they consider a stationary-phase
approximation only for the periodic (real time) solutions. We will
show in Sec.~\ref{sec:bounce}, our calculations for the periodic
classical solutions (after analytic continuation) reproduce their
result.

% static solution does not contribute
It is amazing that ``static solutions'' such as the false vacuum
solution $\xcl(\tau)=\xF$ do not contribute to the free energy, since
such a solution exists only at a single value of $y=\xF$ so that it
has measure zero in the $y$ integration.

% bounce and shifted bounce
The periodic solutions are (finite-$T$) \textit{bounce} and
\textit{shifted bounce} solutions, which are finite-temperature analogs
of the zero-temperature bounce solution. The (usual) bounce solution
defined at zero temperature starts at the top ($x=\xF$) of the inverted
potential $-V(x)$, moving down and bouncing at $x=\xR \ (V(\xR) =
V(\xF))$, and going back to the top. At finite temperature, because it
travels during only the finite period $\beta \hbar$, a finite-$T$ bounce
starts in the middle of the slope ($x=\xb$) of the inverted potential,
with velocity zero, bounces at $x=\xr \ (V(\xr)=V(\xb))$, and goes back
to the original point. See Fig.~\ref{fig:three_solutions}.

% no bounce sol. with the period shorter than \beta_0\hbar
It immediately follows that such a finite-$T$ bounce has a minimum
period: as the period decreases, the corresponding finite-$T$ bounce
moves only in the vicinity of $x_0$, the top of the potential
wall. Since the inverted potential there may be well approximated as
$-V(x) \sim \half |V''(x_0)|(x-x_0)^2$, the upper limit of the period is
given by $\beta_0\hbar \equiv 2\pi/|V''(x_0)|$. There is no bounce
solution which has the period smaller than $\beta_0\hbar$. In this
paper, we restrict ourselves to the temperature less than
$\beta_0^{-1}$. Above it, the main cause of the decay would be thermal
rather than quantum-mechanical tunneling.

\begin{figure}[h]
 \includegraphics[width=0.49\linewidth,clip]{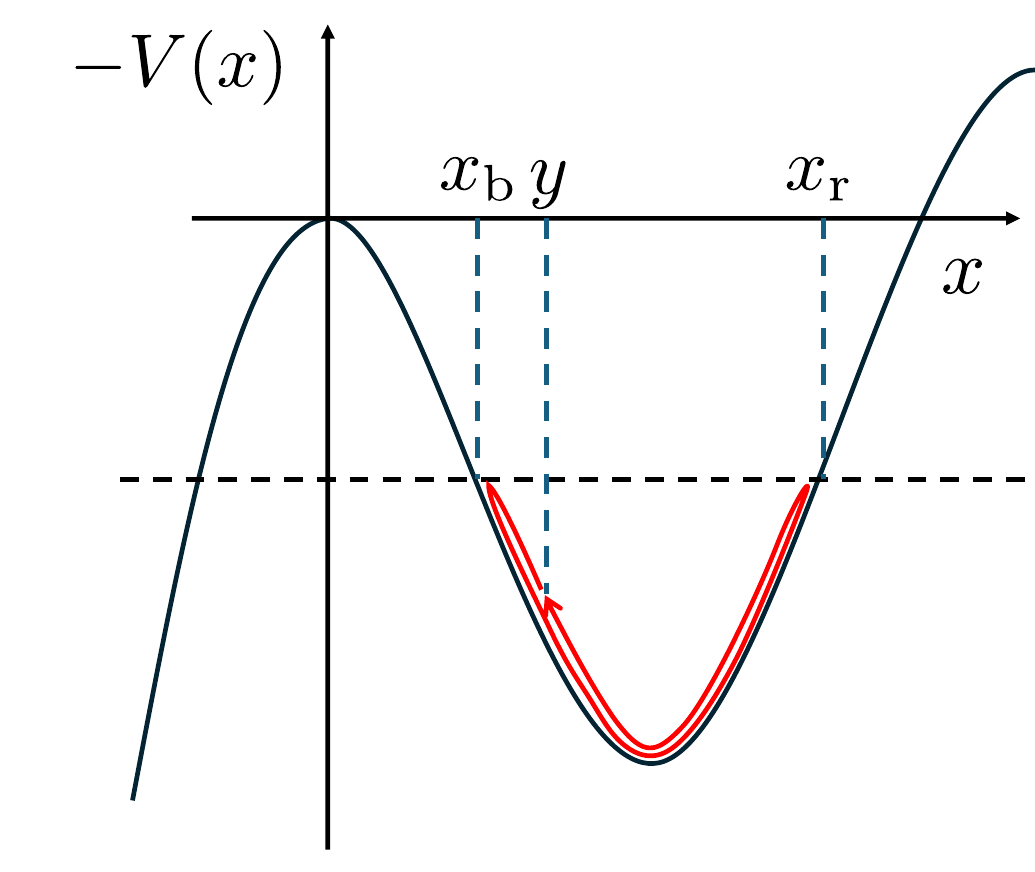}
 \includegraphics[width=0.49\linewidth,clip]{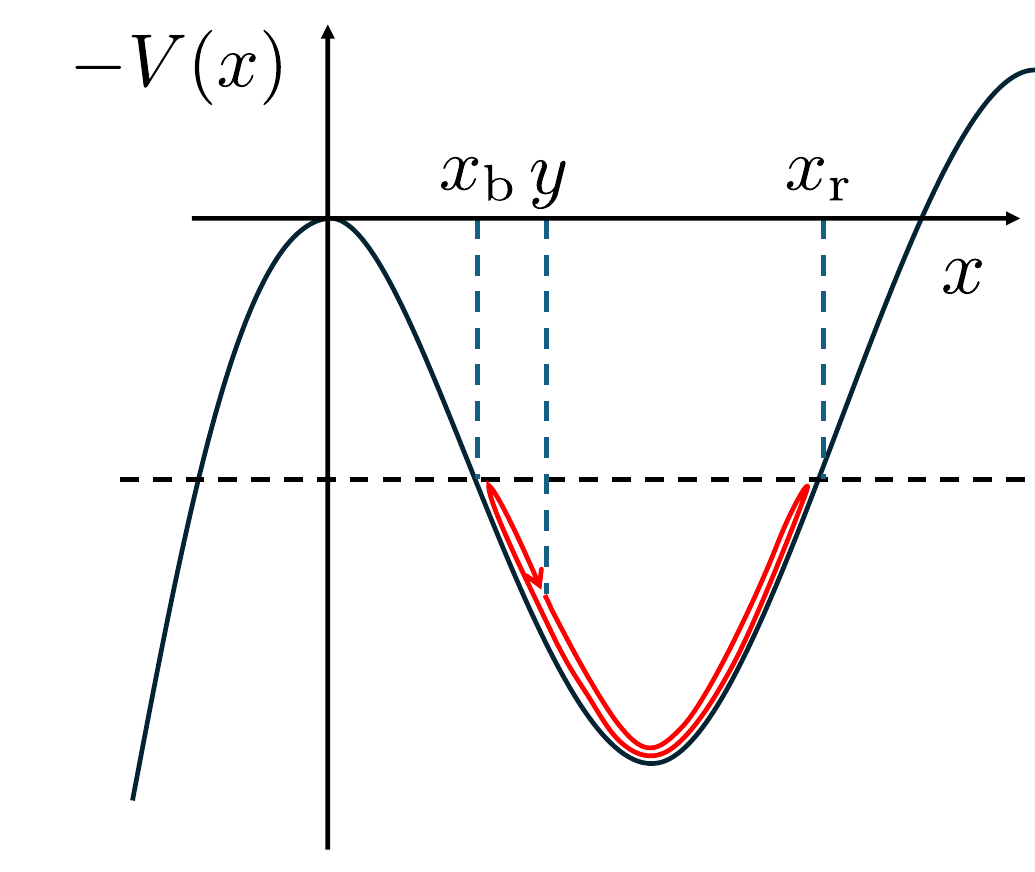}
 \caption{\label{fig:shifted_bounce}
 Shifted bounce solutions. There are two shifted bounce solutions for a
 given energy (or the period) and a starting/ending point $y$.}
\end{figure}

% shifted bounce
Shifted bounce solutions are the time-translated solutions of the
finite-$T$ bounce solution. See Fig.~\ref{fig:shifted_bounce}. At zero
temperature, two bounce solutions which are related by time translation
are not discriminated. On the other hand, at finite temperature, the
time-translated solutions have nonzero initial velocity so that it is
useful to distinguish them from the finite-$T$ bounce solution.

% shot solutions
The ``just-returning-back'' solutions are called shot
solutions~\cite{PhysRevD.95.085011}. There are two kinds of shot
solutions. At low temperature, shot solutions spend most of their time
either near the false vacuum point $\xF$ (shot-F) or near the true
vacuum point $\xT$ (shot-T). At zero temperature limit, the shot-F
solution behaves like the (static) false vacuum solution.  Perhaps
surprisingly, the shot solutions play a crucial role at finite
temperature, though they have not attracted much attention before.

% choice of saddle points and the steepest decent contour
If we include all of these saddle-points contributions in the partition
function~\eqref{partition_function}, the result would be real, as it
looks like. It is the choice of the saddle points and the steepest
descent contour that leads to the imaginary part. The choice of the
steepest descent contour corresponds to the choice of the boundary
condition. (Remember a similar situation in the derivation of the WKB
connection formula by the saddle-point method in elementary quantum
mechanics.)  What we need for calculating the decay rate is the steepest
descent contour which corresponds to the
Gamow-Siegert~\cite{Gamow:1928zz, Siegert:1939zz} (GS) boundary
condition, that is, there are only ``outgoing (decaying) waves''.

% shot-T solutions are irrelevant
It is easy to infer that the shot-T solutions are not relevant to the
false-vacuum decay, since they are related to the true vacuum outside
the potential wall. The inclusion of the shot-T solutions in the
saddle-point approximation would correspond to the inclusion of the
``incoming waves'' and the information on the true vacuum.

% shot-F solutions are relevant
On the other hand, the shot-F solutions are related to the false vacuum,
and the finite-$T$ bounce and shifted bounce solutions are directly
related to the tunneling of the potential wall. Note that, at zero
temperature, the contributions from the false vacuum and the bounce
saddle points lead to the imaginary part of the ground-state energy, the
(minus half of the) decay rate~\cite{PhysRevD.95.085011}.

% the importance of the shot solutions
There are two shot solutions with $y$ just outside the regions $[\xb,
\xr]$, with the same period. They have different values of classical
action, and more importantly, one of them supplies the imaginary part to
the free energy!

\begin{figure}[h]
 \includegraphics[width=\linewidth,clip]{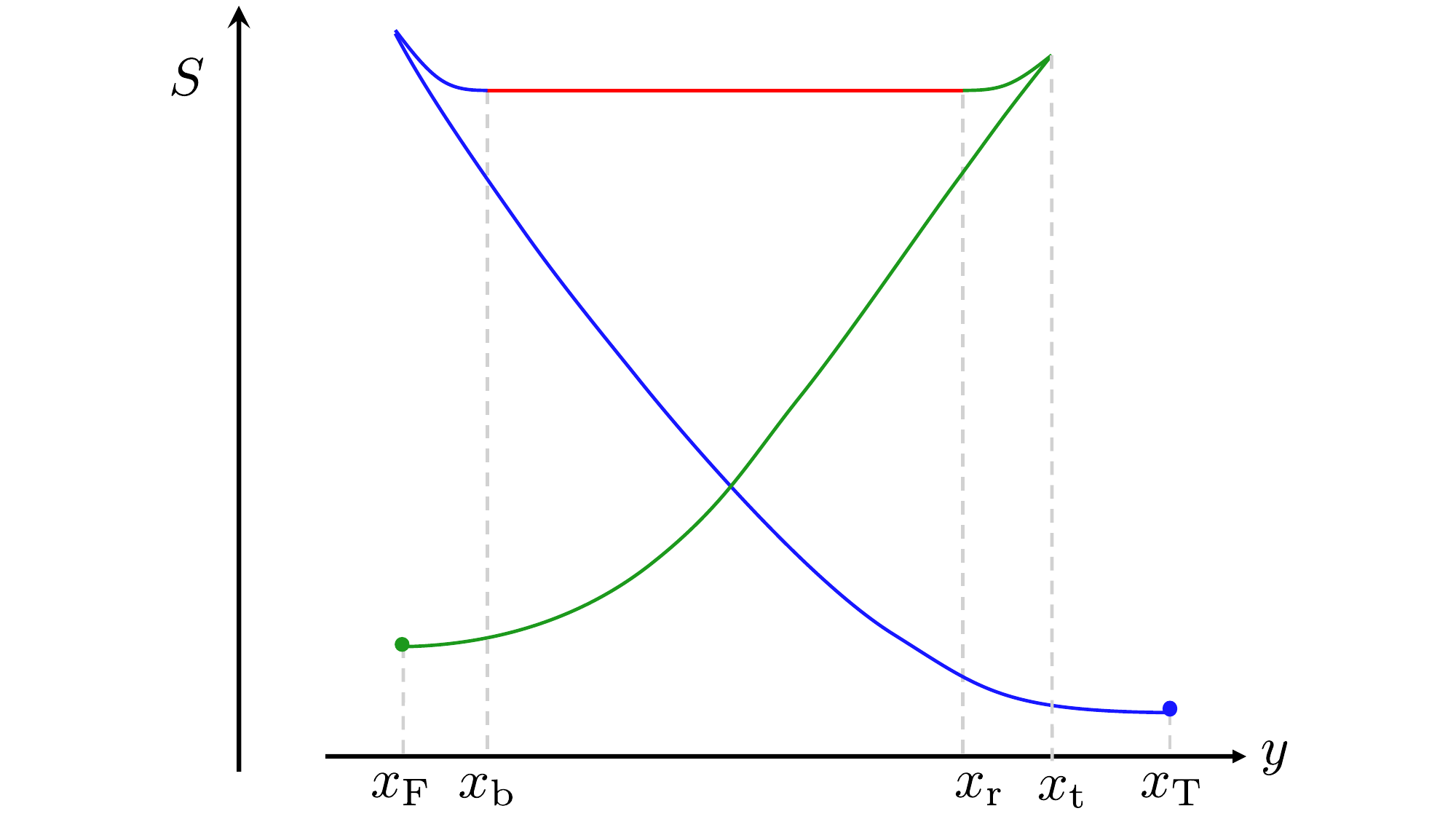}
 \caption{\label{fig:three_acitons} A schematic presentation for the
 actions of the classical solutions for various values of $y$. The
 horizontal straight red line corresponds to the finite-$T$ bounce and
 shifted bounce solutions. The green lines show the actions of shot-F
 solutions. For $y\in (\xr, \xt)$, there are two shot-F
 solutions. Similarly, the actions of shot-T solutions are represented
 as blue lines.}
\end{figure}

% the imaginary and the real branches
Fig.~\ref{fig:three_acitons} shows that finite-$T$ bounce, shifted
bounce, shot-F, and shot-T solutions are intimately connected and form
three branches.  Ignoring the shot-T solutions (because they are
irrelevant to the decay of the false vacuum), there are two solutions
for each value of $y$ in the region $[\xb, \xt]$, where $\xt$ is the
largest value of $y$ for which the shot-F solutions exist. We call the
set of solutions for which the action is larger the \textit{imaginary}
branch, consisting of the finite-$T$ bounce and the shifted bounce
solutions together with one of the shot-F solutions that supplies the
imaginary part to the free energy. The other is called the \textit{real}
branch, consisting of the shot-F solutions that give purely real
contributions to the free energy.

% Gamow-Siegert partition function
Thus, at finite temperature, we calculate the contributions to the
partition function from the finite-$T$ bounce and shifted bounce, $\ZB$,
and from the shot-F, $\ZsF$, to obtain
\begin{equation}
 Z_{GS} = \ZsF^{(\text{R})} +\half \left(Z_B+\ZsF^{(\text{I})}\right),
\end{equation}
where $\ZsF^{(\text{R})}$ and $\ZsF^{(\text{I})}$ are contribution from
the shot-F solutions in the real and imaginary branch, respectively.
The reason why the factor $1/2$ appears in the second term is explained
by Callan and Coleman~\cite{PhysRevD.16.1762}. 
%%
%{\color{red}%%
In their case, the corresponding partition function is
\begin{equation}
 Z_{CC} = Z_\text{FV} +\half Z_{\text{B}},
\end{equation}
where $Z_{\text{F}}$ is the contribution of the static false vacuum
solution $x_{\text{cl}}(\tau) = \xF$. The shot-F solution plays a role
of the static false vacuum solution in our case.
%
%}

% decay rate and comparison with the one by Affleck
As we have explained, $Z_{GS}$ is complex; the decay rate may be given
as
\begin{equation}
 \Gamma = -\frac{2}{\hbar} \text{Im} (F_{GS}) 
  = \frac{2}{\beta\hbar} \tan^{-1}
  \left(\frac{\text{Im} Z_{GS}}{\text{Re} Z_{GS}}\right).
\end{equation}
This formula is quite different from the one given by Affleck. We will
numerically compare the temperature dependence of the decay rates
calculated with two formulae in Sec.~\ref{sec:num}. 

% structure of the paper
The rest of the paper is structured as follows: In
Sec.~\ref{sec:bounce}, we calculate the finite-$T$ bounce and the
shifted bounce contributions, $\ZB$. Because a shifted bounce has a
nonzero initial velocity, its time-derivative of the solution does not
satisfy the Dirichlet boundary condition. Thus the usual collective
coordinate method does not work. The trace integral over $y$ in
Eq.~\eqref{partition_function} plays a similar role to the collective
coordinate integral. The contributions from the shot-F saddle points,
$\ZsF$, will be discussed in Sec.~\ref{sec:shot}. We then complete the
decay rate calculation. In Sec.~\ref{sec:zero}, we first re-examine the
collective coordinate treatment of the zero mode by Callan and
Coleman~\cite{PhysRevD.16.1762} If one takes the finite time-interval
``regularization'' seriously, the Dirichlet boundary conditions which
they work with is inconsistent with the collective coordinate treatment
of the zero mode. The collective coordinate method works only with the
periodic boundary conditions. We then calculate the zero-temperature
limit of our result.  It is different from that of Callan and
Coleman. In Sec.~\ref{sec:num}, we perform some numerical calculations
to see the physical implications of our formula. We compare the
zero-temperature limit of our result with Callan-Coleman result as well
as with the WKB result given in Ref.~\cite{PhysRevD.95.085011}, changing
the height of the potential barrier. We find that our formula gives a
similar result to that of the WKB formula, rather than to that of Callan
and Coleman. We then show the temperature dependence of the decay rate
and compare it with the result of Affleck~\cite{PhysRevLett.46.388}.  As
naively expected, our decay rate increases with the increasing
temperature. The Affleck's formula, on the other hand, leads to a
perplexing result in certain cases: the decay rate \textit{decreases}
with the increasing temperature to a certain point and then
increases. We summarize our result in Sec.~\ref{sec:summary}.

% appendices
In the following sections, we employ dimensionless variables for
notational simplicity and numerical calculations. In
Appendix~\ref{app:dim}, we explain how to introduce them. Superficially,
it corresponds to the use of the unit in which $\hbar=1$ and $m=1$. In
Appendix~\ref{app:det}, we illustrate the method of calculating the
ratio of determinants defined with the periodic boundary conditions, and
show that it becomes equal to the ratio defined with the Dirichlet
boundary conditions in the limit of the infinite interval.

\section{finite-$T$ bounce and shifted bounce solutions}
\label{sec:bounce}

% finite-T bounce
Let us denote the finite-$T$ bounce solution as $\xB(\tau)$, which
satisfies the classical equation of motion~\eqref{EOM}, with the
boundary conditions;
\begin{equation}
 \xB(\pm \beta/2) = \xb,
  \quad
  \dotxB(\pm \beta/2) =0.
\end{equation}
It is a periodic solution, satisfying
\begin{equation}
 \xB(\tau+\beta) = \xB(\tau)
\end{equation}
for an arbitrary value of $\tau$. We denote the value of the classical
action for the finite-$T$ bounce solution as $\SB$.  The finite-$T$
bounce solution has the (Euclidean) energy $E$,
\begin{equation}
 E= \half \dotxB^2-V(\xB),
  \label{bounce_E}
\end{equation}
which is equal to $-V(\xb)$. The energy is determined as a function of
$\beta$, or vice versa.

% saddle-point approx. for the finite-T bounce
The path integral
\begin{equation}
 \zb \approx \int_{x(\pm \beta/2)=\xb}\mathcal{D}x(\tau)\; e^{-S[x]}
\end{equation}
may be evaluated by the saddle-point approximation. By substituting
Eq.~\eqref{def_r}, we have
\begin{equation}
 \zb \approx  e^{-\SB} \int_{r(\pm \beta/2)=0}
  \mathcal{D}r(\tau)
  e^{-\Delta \SB[r]},
\end{equation}
where
\begin{equation}
 \Delta \SB[r]
  = \half \int_{-\beta/2}^{\beta/2}\! d\tau\; r(\tau)
  \left(-\frac{d^2}{d\tau^2}+V''(\xB(\tau))\right)r(\tau).
\end{equation}
The Gaussian integration leads to the determinant of the ``fluctuation
operator''. But, as is well known, there are two things to be taken into
account.

% zero mode and negative mode
The first is the existence of the zero mode. Actually, by
differentiating Eq.~\eqref{EOM} with respect to $\tau$, we obtain
\begin{equation}
 \left(-\frac{d^2}{d\tau^2}+V''(\xB(\tau))\right)
  \dotxB(\tau)=0.
\end{equation}
Since $\dotxB$ satisfies the boundary conditions $\dotxB(\pm
\beta/2)=0$, it may be regarded as a part of $r(\tau)$. We need to omit
the zero eigenvalue from the determinant.

The second, since the zero mode $\dotxB$ has a node, there must be a
negative mode. The existence of the negative mode causes the imaginary
part in the path integral, and leads eventually to the decay rate of the
false vacuum. See Ref.~\cite{PhysRevD.16.1762, Coleman_1985,
Coleman:1987rm}.

% difference between T=0 and T\ne0
% shifted bounces
At zero temperature, a shift of the (Euclidean) time for a bounce
solution is related to the zero mode. (See however Sec.~\ref{sec:CC} for
the issues of the introducing a finite time interval.) At finite
temperature, on the other hand, the story is not that simple. The
shifted bounce solution
\begin{equation}
 \xsB(\tau;\tau_c) \equiv \xB(\tau+\tau_c) \quad
  (0<\tau_c <\beta),
\end{equation}
does not satisfy $\dotxsB(\pm \beta/2; \tau_c)=0$ so that it is not
regarded as a part of $r(\tau)$. (Remember the boundary conditions
Eq.~\eqref{r_bc}.) There is no zero mode in the Gaussian integration.

% shifted bounce contributions
We consider the sum of the saddle-point contributions of the shifted
bounce solutions,
\begin{equation}
 \ZB \approx \mathcal{N} e^{-\SB}
  \int_{\xb}^{\xr} dy \int_{r(\pm \beta/2)=0}\!\!\!
  \mathcal{D}r(\tau)\; e^{-\Delta \SsB[r;y]},
\end{equation}
where
\begin{equation}
 y=\xsB(\pm \beta/2; \tau_c)= \xB(\pm \beta/2+\tau_c)
  \label{ytotauc}
\end{equation}
is the starting and ending point, and
\begin{align}
 &\Delta \SsB[r; y]
 \notag \\
 =&
  \half \int_{-\beta/2}^{\beta/2}\! d\tau\; r(\tau)
  \left(-\frac{d^2}{d\tau^2}+V''(\xsB(\tau;\tau_c))\right)r(\tau).
 \label{DeltaS_sB}
\end{align}
The normalization constant $\mathcal{N}$ will be fixed later.

% only y in [x_b,x_r]
Note that the shifted bounce solutions exit only for $y$ between $\xb$
and $\xr$ and the value of the classical action for a shifted bounce
solution is the same as that for the finite-$T$ bounce solution for any
value of $y$.

% two shifted bounce solutions for a single value of y
There are actually two shifted bounce solutions for any single value of
$y \in (\xb, \xr)$; one moves initially to the right and the other to
the left, as shown in Fig.~\ref{fig:shifted_bounce}. Thus the
expression~\eqref{DeltaS_sB} should be understood as an abbreviated one;
it is the sum of these two contributions. These two solutions are related
by the change of $\tau_c$ to $\beta-\tau_c$.

% combined by changing y to \tau_c
These two contributions can be neatly combined by changing variable from
$y$ to $\tau_c$ through $y=\xsB(\beta/2;\tau_c)$,
\begin{align}
 \ZB &\approx e^{-\SB}\int_{0}^{\beta}
 d\tau_c\left|\dotxsB(\beta/2;\tau_c)\right|
 \notag \\
 &\times
 \frac{\mathcal{N}}
 {\sqrt{\det\left[-\frac{d^2}{d\tau^2}+V''(\xB(\tau+\tau_c))\right]}}.
 % \int_{r(\pm \beta/2)=0} \!\!\!\mathcal{D}r(\tau) \;
 % e^{-\Delta S_{\text{sB}}[r;\tau_c]}
 \label{ZB}
\end{align}

% Gel'fand \& Yaglom
We invoke the Gel'fand-Yaglom theorem~\cite{Gelfand:1959nq} (see also
Appendix~1 of Ref.~\cite{Coleman_1985}) to
evaluate the determinant. According to the theorem, the ratio of the
two determinant can be written as the ratio of the boundary value of
the eigenfunctions,
\begin{equation}
 \frac{\det\left[-\frac{d^2}{d\tau^2}+V''(\xB(\tau+\tau_c))\right]}
  {\det\mathcal{O}}
  =\frac{\psi(\beta/2)}{\psi_{\mathcal{O}}(\beta/2)},
  \label{ratio_det}
\end{equation}
where $\psi(\tau)$ is the eigenfunction of the fluctuation operator,
\begin{equation}
 \left[
  -\frac{d^2}{d\tau^2}+V''(\xB(\tau+\tau_c))
 \right]\psi(\tau)=0,
 \label{eigen_eq_sB}
\end{equation}
$\mathcal{O}$ is a reference operator of the form
\begin{equation}
 \mathcal{O} =-\frac{d^2}{d\tau^2}+W,
\end{equation}
and $\psi_{\mathcal{O}}(\tau)$ is the eigenfunction of the reference
operator,
\begin{equation}
 \mathcal{O}\psi_{\mathcal{O}}(\tau)=0.
\end{equation}
These eigenfunctions are supposed to satisfy the following boundary
conditions;
\begin{align}
 \psi(-\beta/2) &=\psi_{\mathcal{O}}(-\beta/2) =0, \\
 \dot{\psi}(-\beta/2)&= \dot{\psi}_{\mathcal{O}}(-\beta/2)=1.
\end{align}
We assume that the operator $\mathcal{O}$ and its eigenfunction
$\psi_{\mathcal{O}}(\tau)$ is known.

% another eigenfunction \nu
Since $\mu(\tau)\equiv \dotxsB(\tau;\tau_c)$ does not satisfy the
boundary conditions mentioned above, we need another eigenfunction 
with the eigenvalue zero to construct $\psi(\tau)$. It is provided by the
derivative of $\xsB(\tau;\tau_c)$ with respect to the energy. Actually,
by differentiating the equation of motion
\begin{equation}
 -\frac{d^2}{d\tau^2}\xsB(\tau;\tau_c)+V'(\xsB(\tau;\tau_c))=0
\end{equation}
with respect to $E$, we obtain
\begin{equation}
  \left[
  -\frac{d^2}{d\tau^2}+V''(\xB(\tau+\tau_c))
 \right]\der_E\xsB(\tau;\tau_c)=0,
\end{equation}
which shows that $\nu(\tau)\equiv \der_{E}\xsB(\tau;\tau_c)$ is the one
we need~\cite{Marino:2015yie}.

% construction of \psi
The eigenfunction $\psi(\tau)$ may be written as a linear combination of
$\mu(\tau)$ and $\nu(\tau)$;
\begin{equation}
 \psi(\tau)= A \mu(\tau) + B\nu(\tau).
\end{equation}
By imposing the boundary conditions, we get
\begin{align}
 &A\mu(-\beta/2)+B\nu(-\beta/2)=0,\label{psi_bc1} \\
 &A\dot{\mu}(-\beta/2)+B\dot{\nu}(-\beta/2)=1. \label{psi_bc2}
\end{align}
The constants $A$ and $B$ are easily obtained by noting that the
Wronskian
\begin{equation}
 W\equiv \mu(\tau)\dot{\nu}(\tau)-\dot{\mu}(\tau)\nu(\tau)
\end{equation}
is equal to $1$. Actually,
\begin{align}
 W &= \dotxsB\der_E\dotxsB-\ddotxsB\der_E\xsB \notag \\
 &= \der_E\left[\half \dotxsB^2-V(\xsB)\right]=1.
\end{align}
Because $W$ is a constant, it may be evaluated at $\tau=-\beta/2$.  We
obtain $A=-\nu(-\beta/2)$ and $B=\mu(-\beta/2)$.

% more useful expression
Thus, the ratio of the determinants Eq.~\eqref{ratio_det} has been
calculated in terms of $\mu$ and $\nu$. We can now rewrite the result
in a more useful way.

% using periodicity
Because $\xsB(\tau; \tau_c)$ is periodic, $\xsB(\tau+\beta;\tau_c) =
\xsB(\tau; \tau_c)$, we obtain
\begin{equation}
 \nu(\tau+\beta) +\mu(\tau+\beta)\frac{d\beta}{dE} =\nu(\tau),
\end{equation}
by differentiating it with respect to $E$. By setting $\tau=-\beta/2$,
we get
\begin{equation}
 \nu(\beta/2)-\nu(-\beta/2) = -\mu(\beta/2)\frac{d\beta}{dE}.
\end{equation}
By using this identity and the periodicity of $\mu$, we rewrite
$\psi(\beta/2)$ as
\begin{align}
 \psi(\beta/2)&=-\nu(-\beta/2)\mu(\beta/2)+\mu(-\beta/2)\nu(\beta/2)
 \notag \\
 &= \mu(-\beta/2)\left(\nu(\beta/2)-\nu(-\beta/2)\right)
 \notag \\
 &=-\left[\mu(-\beta/2)\right]^2\frac{d\beta}{dE}.
 \label{psi_betahalf}
\end{align}
Because the period $\beta$ is a increasing function of $E$,
$d\beta/dE>0$, and $\mu(-\beta/2)\ne0$, we have $\psi(\beta/2)<0$. It
means that the fluctuation operator for $\xsB$ does not contain a zero
mode, but an odd number of negative modes. 

% reference fluctuation operator
% removal of the zero eigenvalue
We want to take the fluctuation operator for the finite-$T$ bounce
solution as the reference operator;
$\mathcal{O} = -d^2/d\tau^2 + V''(\xB(\tau))$. The corresponding
eigenfunction that satisfies the boundary conditions at $\tau = -\beta/2$
is $\psi_{\mathcal{O}}(\tau) = \dotxB(\tau)/\ddotxB(-\beta/2)$. As we
mentioned earlier, however, there is a zero mode for the operator and
$\psi_{\mathcal{O}}$ is actually the zero mode because it satisfies
$\psi_{\mathcal{O}}(\beta/2)=0$. 
To remove the zero eigenvalue from the determinant, we introduce an
arbitrary number $\lambda$ and consider the determinant,
$\det\left[\mathcal{O}-\lambda \right]$. Instead of zero, the operator
$\mathcal{O}-\lambda$ has the eigenvalue $-\lambda$. We define the
zero eigenvalue removed determinant ${\det}'\mathcal{O}$ as
\begin{equation}
 {\det}'\mathcal{O} 
=\lim_{\lambda\to0}
\frac{\det\left[\mathcal{O}-\lambda\right]}{-\lambda}.
\end{equation}
We consider $\psi_{\mathcal{O}\lambda}$ which satisfies
\begin{equation}
 \left[\mathcal{O}-\lambda\right]\psi_{\mathcal{O}\lambda}(\tau)=0,
  \label{Ominuslambda}
\end{equation}
with the boundary conditions $\psi_{\mathcal{O}\lambda}(-\beta/2)=0$ and
$\dot{\psi}_{\mathcal{O}\lambda}(-\beta/2)=1$. Multiplying
$\dotxB(\tau)$ with Eq.~\eqref{Ominuslambda} and
integrating over $\tau$, we get
\begin{align}
 0&= \int_{-\beta/2}^{\beta/2} d\tau \;
 \dotxB(\tau)\left[\mathcal{O}-\lambda\right]\psi_{\mathcal{O}\lambda}(\tau)
 \notag \\
 &= \ddotxB(\beta/2)\psi_{\mathcal{O}\lambda}(\beta/2)-\lambda
 \int_{-\beta/2}^{\beta/2} d\tau \;
 \dotxB(\tau)\psi_{\mathcal{O}\lambda}(\tau),
 \label{to_get_lambda}
\end{align}
by integrating by parts and using the boundary conditions. We obtain
\begin{equation}
 \frac{\psi_{\mathcal{O}\lambda}(\beta/2)}{-\lambda} = 
  -\frac{\int_{-\beta/2}^{\beta/2} d\tau\;
  \dotxB(\tau)\psi_{\mathcal{O}\lambda}(\tau)}
  {\ddotxB(\beta/2)}.
\end{equation}

% combining
Combining all of them, we have
\begin{align}
 &\frac{\det\left[-\frac{d^2}{d\tau^2}+V''(\xB(\tau+\tau_c))\right]}
 {{\det}'\mathcal{O}} 
 \notag \\
 &= \lim_{\lambda\to0}
 \frac{\det\left[-\frac{d^2}{d\tau^2}+V''(\xB(\tau+\tau_c))\right]}
 {\det\left[\mathcal{O}-\lambda\right]/(-\lambda)}
 \notag \\
 &= \lim_{\lambda\to0}
 \frac{\psi(\beta/2)}{\psi_{\mathcal{O}\lambda}(\beta/2)/(-\lambda)}
 \notag \\
 &= - \lim_{\lambda\to0}
 \frac{\ddotxB(\beta/2)}
 {\int_{-\beta/2}^{\beta/2}d\tau\; \dotxB(\tau)\psi_{\mathcal{O}\lambda}(\tau)}
 \psi(\beta/2)
 \notag \\
 &=-\frac{\ddotxB(-\beta/2)\ddotxB(\beta/2)}
 {\left|\left|\dotxB\right|\right|^2}\psi(\beta/2),
\end{align}
where we have used $\lim_{\lambda\to0}\psi_{\mathcal{O}\lambda}(\tau)
=\psi_{\mathcal{O}}(\tau) = 
\dotxB(\tau)/\ddotxB(-\beta/2)$, and have introduced the notation
\begin{equation}
 \left|\left|\dotxB\right|\right|^2 =
  \int_{-\beta/2}^{\beta/2}d\tau\;
  \left(\dotxB(\tau)\right)^2.
\end{equation}

% Z_B
Finally the contribution of the shifted bounce solution~\eqref{ZB}
can be written as
\begin{align}
 \ZB&\approx \mathcal{N}\frac{\beta e^{-\SB}}
  {\sqrt{{\det}'\left[-\frac{d^2}{d\tau^2}+V''(\xB(\tau))\right]}}
 \notag \\
 &\times\sqrt{\frac{\left|\left|\dotxB\right|\right|^2}
 {\ddotxB(-\beta/2)\ddotxB(\beta/2)}}
 \left[\frac{d\beta}{dE}\right]^{-\half}
 \notag \\
 &=\mathcal{N}\frac{i\beta e^{-\SB} }
 {\sqrt{\left|{\det}'\left[-\frac{d^2}{d\tau^2}+V''(\xB(\tau))\right]\right|}}
 \frac{\left|\left|\dotxB\right|\right|}{\left|\ddotxB(\beta/2)\right|}
\left[\frac{d\beta}{dE}\right]^{-\half}.
 \label{ZBwithN}
\end{align}
Note that the Jacobian in Eq.~\eqref{ZB} cancels the
$\left[\mu(-\beta/2)\right]^2$ in Eq.~\eqref{psi_betahalf} and the
factor $i$ appears because the ${\det}'\mathcal{O}<0$. The integration
over $\tau_c$ is trivial because the integrand is now independent of
$\tau_c$. 
% By the definitions of $\SB$ and $E$, we may write 
% \begin{equation}
%  \left|\left|\dotxB\right|\right|= \sqrt{\SB+\beta E}.
% \end{equation}

% normalization
The normalization constant $\mathcal{N}$ may be fixed by using the
harmonic oscillator potential~\cite{PhysRevD.16.1762,Coleman_1985},
\begin{equation}
 \frac{\mathcal{N}}{\sqrt{\det\left[-\frac{d^2}{d\tau^2}+\omega^2\right]}}
  =\sqrt{\frac{\omega}{2\pi\sinh(\omega\beta)}}.
  \label{N_harmonic}
\end{equation}
Furthermore, by using Gel'fand-Yaglom theorem once again, we have
\begin{align}
 \frac{\det\left[-\frac{d^2}{d\tau^2}+\omega^2\right]}
  {{\det}'\left[-\frac{d^2}{d\tau^2}+V''(\xB(\tau))\right]}
 &=
 -\frac{\left|\ddotxB(\beta/2)\right|^2}{\left|\left|\dotxB\right|\right|^2}
 \frac{\sinh(\omega\beta)}{\omega}.
 \label{ratio_of_dets}
\end{align}
Note that $\psi_{\text{HO}}(\tau)\equiv
(1/\omega)\sinh(\omega(\tau+\beta/2))$ is an eigenfunction of the
operator $-d^2/d\tau^2+\omega^2$ with eigenvalue zero, and satisfies the
boundary conditions
$\psi_{\text{HO}}(-\beta/2)=0$ and $\dot{\psi}_{\text{HO}}(-\beta/2)=1$.

% final result for Z_B
We may now rewrite Eq.~\eqref{ZBwithN} in a much simpler form;
\begin{equation}
 \ZB \approx \frac{i\beta}{\sqrt{2\pi}} e^{-\SB} 
  \left[\frac{d\beta}{dE}\right]^{-\half}.
  \label{ZB_final}
\end{equation}

% comparison with DHN
At this point, it is instructive to compare $\ZB$  with Eq.~(2.12) of
Ref.~\cite{Dashen:1974ci},
\begin{equation}
 \tr\; e^{-iHT} = \left(\frac{i}{2\pi\hbar}\right)^{\half}
  T\left|\frac{dE_{\text{cl}}}{dT}\right|^{\half}
  e^{i\left(S_{\text{cl}}(T/\hbar)-\pi\right)},
  \label{DHNresult}
\end{equation}
where $T$ is a time interval, or the period of the classical solution.
Changing $T\to -i\beta$, $E_{\text{cl}}\to -E$, and $S_{\text{cl}}\to -i
\SB$ formally, we see that it is identical to our result.

% multi-bounce solutions ignored
The $n$-cycle finite-$T$ and shifted bounce solutions with period
$\beta/n$ are also classical solutions with period $\beta$. We ignore
all of these ``multiple-bounce'' solutions, because (i) we assume that
the $n=1$ contribution is already very small, and (ii) there is a
maximum value of $n$ due to the condition $\beta\ge\beta_0$.

\section{shot solutions and decay rate}
\label{sec:shot}

% description of the shot-F solution
A shot-F solution $\xsF(\tau;y)$ starts at $x=y$ going toward to
$x=\xF$, reaches the highest point $x=\xs$, and returns back to the
original position after the time lapse $\beta$;
\begin{align}
 \xsF(\pm \beta/2;y)&=y,
 \notag \\
 \xsF(0; y)&= \xs,
 \notag \\
 \dotxsF(-\beta/2;y)&=-\dotxsF(\beta/2;y).
  \label{sF_bc}
\end{align}
We denote the corresponding classical action as $\SsF(y)$.

% saddle-point approx.
The saddle-point approximation of the contribution of the shot-F
solutions may be written as
\begin{equation}
 \ZsF \approx\int^{\xt}_{\xb} dy 
  \frac{\mathcal{N} e^{-\SsF(y)}}
  {\sqrt{\det \left[-\frac{d^2}{d\tau^2}+V''(\xsF(\tau;y))\right]}}.
\end{equation}
The upper limit $\xt$ will be discussed shortly.

% no zero mode
Note that, since the initial and final velocities are nonzero,
$\dotxsF(\tau)$ is not a zero mode of the fluctuation operator, in
contrast to the bounce case.

% Gel'fand-Yaglom
By using Eq.~\eqref{N_harmonic} and Gel'fand-Yaglom theorem, we obtain
\begin{equation}
 \ZsF\approx \frac{1}{\sqrt{2\pi}}
  \int dy \frac{e^{-\SsF(y)}}{\sqrt{\psi_{\text{sF}}(\beta/2;y)}},
\end{equation}
where $\psi_{\text{sF}}(\tau;y)$ is the eigenfunction
\begin{equation}
 \left[-\frac{d^2}{d\tau^2}+V''(\xsF(\tau; y))\right]
  \psi_{\text{sF}}(\tau;y) = 0,
\end{equation}
with the boundary conditions $\psi_{\text{sF}}(-\beta/2; y)=0$ and
$\dot{\psi}_{\text{sF}}(-\beta/2; y)=1$.

% eigenfunction
We can construct this eigenfunction by using
 $\mu_{\text{s}}(\tau;y)\equiv \dotxsF(\tau;y)$ and
 $\nu_{\text{s}}(\tau;y)\equiv \der_E\xsF(\tau;y)$ in a similar manner
 to the case of the shifted bounce solutions;
\begin{equation}
 \psi_{\text{sF}}(\tau;y) 
  = -\nu_{\text{s}}(-\beta/2;y)\mu_{\text{s}}(\tau;y)
  +\mu_{\text{s}}(-\beta/2;y) \nu_{\text{s}}(\tau; y).
\end{equation}
Thus, we have 
\begin{equation}
 \psi_{\text{sF}}(\beta/2;y) = 
  2\mu_{\text{s}}(-\beta/2; y)\nu_{\text{s}}(-\beta/2;y),
  \label{psi_sF}
\end{equation}
by using Eq.~\eqref{sF_bc} and $\nu_{\text{s}}(\beta/2;y) =
\nu_{\text{s}}(-\beta/2;y)$.

% nu_s requires careful examination
For the potential depicted in Fig.~\ref{fig:potential}, the initial
velocity is negative; $\mu_{\text{s}}(-\beta/2;y)<0$. On the other hand,
the determination of the sign of $\nu_{\text{s}}(-\beta/2;y)$ requires
more careful examination.

% naive expectation
If the energy increases, the turning point $\xs$ gets closer to $\xF$,
i.e., shifts to the left. One may naively expect that the starting point
also shifts to the left in order for the particle to return back to it
after the time lapse $\beta$, and thus
$\nu_{\text{s}}(-\beta/2;y)<0$. If it is the case,
$\psi_{\text{sF}}(\beta/2;y)>0$ so that $\ZsF$ is real.

% naive expectation is not always correct
Amazingly, this expectation is not always correct. We numerically find
that there is a region of $y \in (\xr,\xt)$, for which there are
\textit{two} shot-F solutions for a given time lapse $\beta$, and one of
which has $\nu_{\text{s}}(-\beta/2;y)<0$, and the other
$\nu_{\text{s}}(-\beta/2;y)>0$. The latter shot-F solution leads to an
imaginary partition function, thus may be called $\xsF^{\text{(I)}}$.
The shot-F solution with $\nu_{\text{s}}(-\beta/2;y)<0$ may be called
$\xsF^{\text{(R)}}$.

% dS_{sF}/dE
The eigenfunction~\eqref{psi_sF} may be written in a different way. By
differentiating $\SsF(y)$ with respect to the energy $E$ of the shot-F
solution, we find
\begin{align}
 \frac{d\SsF(y)}{dE} &=\int_{-\beta/2}^{\beta/2} d\tau
  \left[
   \dotxsF\pd{\dotxsF}{E}+V'(\xsF)\pd{\xsF}{E}
  \right]
  \notag \\
 &= \left[\dotxsF\pd{\xsF}{E}\right]_{-\beta/2}^{\beta/2}
 \notag \\
 &\quad
 +\int_{-\beta/2}^{\beta/2} d\tau
 \left[
 -\ddotxsF+V'(\xsF)
 \right]\pd{\xsF}{E}
 \notag \\
 &= -2\mu_{\text{s}}(-\beta/2;y)\nu_{\text{s}}(-\beta/2;y)
 \notag \\
 &= -\psi_{\text{sF}}(\beta/2;y).
\end{align}
For $\xsF^{\text{(R)}}$ it is negative while for $\xsF^{\text{(I)}}$
it is positive.

%\section{Decay rate}

\begin{figure}[h]
 \includegraphics[width=\linewidth,clip]{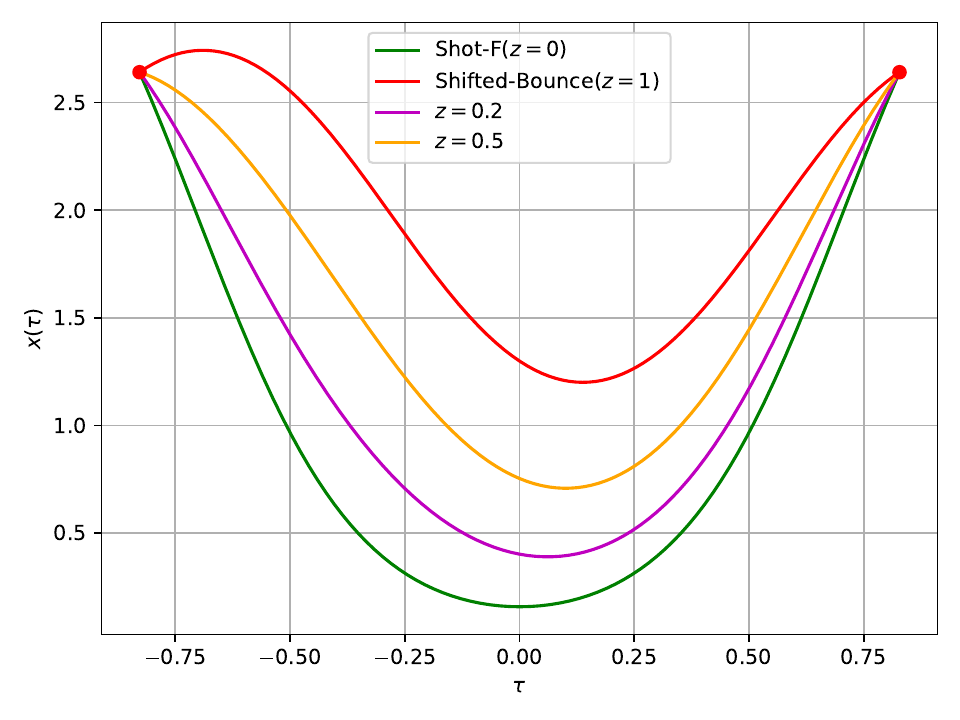}
 \caption{\label{fig:parameterized_paths} A one-parameter family of
 the classical paths $\tilde{x}(\tau;z)$ which interpolates the
 shifted bounce solution ($z=1$) and the shot-F solution ($z=0$),
 numerically calculated for the potential $V(x)=x^4-8x^3+15.9x^2$.}
\end{figure}

% interpolation between shot-F and bounce
It is instructive to construct a one-parameter family of the classical
paths $\tilde{x}(\tau;z)$ which interpolates the (shifted) bounce
solution $\tilde{x}(\tau;1) = \xsB(\tau;\tau_c)$ with $y =
\xsB(\pm\beta/2;\tau_c)$ and the shot-F solution $\tilde{x}(\tau;0) =
\xsF^{\text{(R)}}(\tau;y)$ for each value of $y\in [\xb, \xr]$. We
require that it satisfies (i) $\tilde{x}(\pm \beta/2;z)=y$ and (ii)
$\left[ d\tilde{x}/dz \right]_{z=1} = r_-(\tau;\tau_c)$, where
$r_-(\tau;\tau_c)$ is the eigenfunction of $-d^2/d\tau^2 +
V''(\xsB(\tau;\tau_c))$ with negative eigenvalue.  We use a quadratic
function of $z$,
\begin{align}
 \tilde{x}(\tau;z) &=
 (r_-(\tau;\tau_c)-\xsB(\tau;\tau_c)+\xsF^{(\text{R})}(\tau;y))z^2
 \notag \\
 &\quad +(2\xsB(\tau;\tau_c)-2\xsF^{(\text{R})}(\tau;y)-r_-(\tau;\tau_c))z
 \notag \\
 &\quad +\xsF^{(\text{R})}(\tau;y).
\end{align}
See Fig.~\ref{fig:parameterized_paths}. Remember that a similar
construction was considered by Callan and Coleman to draw Fig.~5 and
Fig.~6 in Ref.~\cite{PhysRevD.16.1762}.

\begin{figure}[h]
 \includegraphics[width=\linewidth,clip]{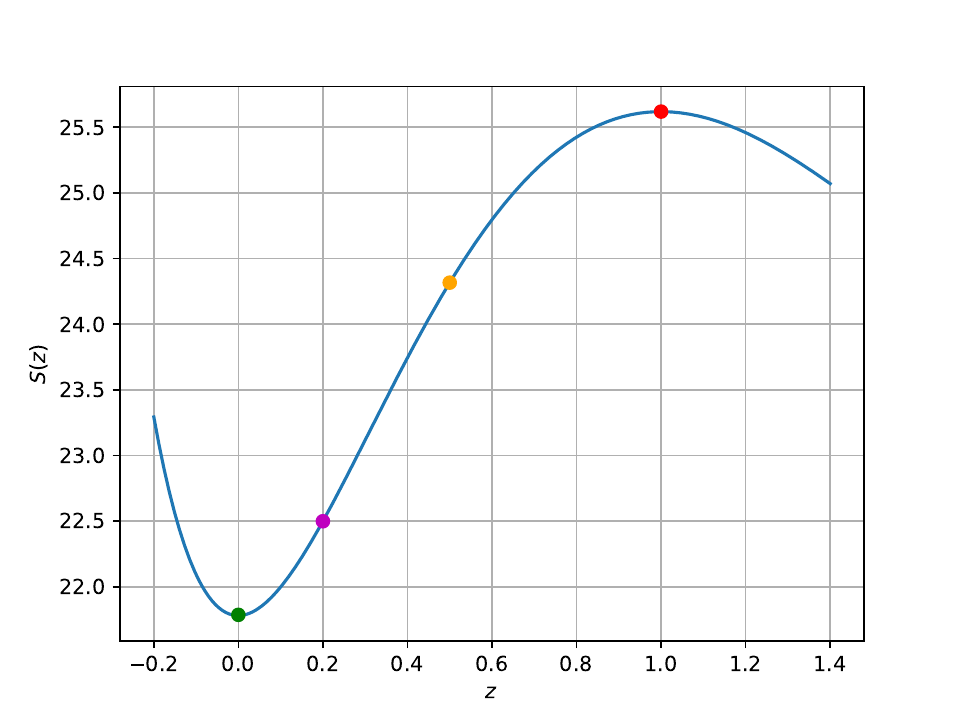}
 \caption{\label{fig:S(z)} The classical action $S(z)$ for the
 parametrized path $\tilde{x}(\tau; z)$. Colored dots correspond to
 the bounce solution (red), the shot-F solution (green), and the
 interpolating paths (purple and orange) given in
 Fig.~\ref{fig:parameterized_paths}. }
\end{figure}

% z=0 and 1: extremum
One can easily show that the action $S(z)\equiv S[\tilde{x}]$ has a minimum
at $z=0$ (shot-F) and a maximum at $z=1$ (bounce). Actually,
\begin{align}
 \frac{dS(z)}{dz} &=\int_{-\beta/2}^{\beta/2} d\tau
 \left[
 \dot{\tilde{x}}\frac{d\dot{\tilde{x}}}{dz}
 + V'(\tilde{x}) \frac{d\tilde{x}}{dz}
 \right]
 \notag \\
 &= \int_{-\beta/2}^{\beta/2} d\tau
 \left[
 -\ddot{\tilde{x}}+V'(\tilde{x})
 \right]\frac{d\tilde{x}}{dz}
\end{align}
shows that it vanishes at $z=0$ and $z=1$, because at these points,
$\tilde{x}$ is a solution of the equation of motion. Furthermore, we
obtain
\begin{align}
 \frac{d^2S(z)}{dz^2}&=\int_{-\beta/2}^{\beta/2} d\tau
 \left[
 -\frac{d\ddot{\tilde{x}}}{dz}+V''(\tilde{x})\frac{d\tilde{x}}{dz}
 \right]\frac{d\tilde{x}}{dz}
 \notag \\
 &\quad+\int_{-\beta/2}^{\beta/2} d\tau
 \left[
 -\ddot{\tilde{x}}+V'(\tilde{x})
 \right]\frac{d^2\tilde{x}}{dz^2}
 \notag \\
 &= \int_{-\beta/2}^{\beta/2} d\tau\;
 \frac{d\tilde{x}}{dz}
 \left[
 -\frac{d^2}{d\tau^2}+V''(\tilde{x})
 \right]\frac{d\tilde{x}}{dz}
 \notag \\
 &\quad+\int_{-\beta/2}^{\beta/2} d\tau
 \left[
 -\ddot{\tilde{x}}+V'(\tilde{x})
 \right]\frac{d^2\tilde{x}}{dz^2},
\end{align}
The second term vanishes at $z=0$ and $z=1$ for the same reason as
before.  For $z=0$, because there is no negative mode nor the zero
mode, the first term is positive; $\left[d^2S(z)/dz^2\right]_{z=0} >
0$. On the other hand, because of the condition (ii), the first term
is negative; $\left[d^2S(z)/dz^2\right]_{z=1} < 0$. $S(z)$ is depicted
in Fig.~\ref{fig:S(z)} for a specific potential.

% shot-F plays a similar role to that of the false-vacuum sol.
It implies that the shot-F solution plays a similar role to that of
the false-vacuum solution $x=\xF$ and the steepest decent contour may
be similar to Callan-Coleman case, and that the contribution of
the bounce solution has the well-known factor $1/2$.

% xsF^(I) plays a similar role to bounce
For $y\in (\xr, \xt)$, the $\xsF^{\text{(I)}}$ plays a similar role
of the finite-$T$ bounce and shifted bounce solutions.

% our decay rate formula at finite temperature
From these considerations, we arrive at the formula for the decay rate
$\Gamma$ at finite temperature,
\begin{align}
 \Gamma &=-2\text{Im} \left(F_{\text{GS}}\right)
  = \frac{2}{\beta} \tan^{-1}
  \left(
 \frac{\text{Im} Z_{\text{GS}}}
 {\text{Re} Z_\text{GS}}
 \right)
 \notag \\
 &=\frac{2}{\beta}\tan^{-1}
  \left(
   \frac{\ZB+\ZsF^\text{(I)}}
   {2\ZsF^\text{(R)}}
  \right),
 \label{our_result}
\end{align}
where
\begin{equation}
 Z_{\text{GS}} = \ZsF^{(\text{R})} 
  +\half \left(\ZB + \ZsF^{(\text{I})}\right),
\end{equation}
with
\begin{align}
 \ZsF^{(\text{R})}
 &\approx \frac{1}{\sqrt{2\pi}}\int_{\xb}^{\xt} dy
 \frac{e^{-S[\xsF^{(\text{R})}(\tau;y)]}}
 {\sqrt{\left|\frac{d\SsF^{(\text{R})}}{dE}\right|}},
 \\
 \ZsF^{(\text{I})}
 &\approx \frac{i}{\sqrt{2\pi}}\int_{\xr}^{\xt} dy
 \frac{e^{-S[\xsF^{(\text{I})}(\tau;y)]}}
 {\sqrt{\left|\frac{d\SsF^{(\text{I})}}{dE}\right|}}.
\end{align}

% numerical solutions will be given in Sec.~\ref{sec:num}
Numerical solutions on the basis of this formula will be discussed in
Sec.~\ref{sec:num}.

\section{zero-temperature limit}
\label{sec:zero}

% In this section we consider the zero-temperature limit for a generic
% potential. The case of symmetric double well potentials, such as
% $V(x)=\lambda (x-\xF)^2(x-\xR)^2 \ (\lambda>0)$, requires special care
% and will be considered in Sec.~\ref{subsec:T-dep}.

\subsection{Zero-temperature result by Callan and Coleman}
\label{sec:CC}

% Callan & Coleman
Before considering the zero-temperature limit of our result, let us
recapitulate the results by Callan and Coleman
(CC)~\cite{PhysRevD.16.1762} for the zero temperature, emphasizing the
finite time interval $T$ (not temperature in this subsection), which is
introduced to regularize the theory. We use the notation of
Ref.~\cite{PhysRevD.16.1762} in this subsection, to make the comparison
easier.

% CC formula
% three points
The CC formula (Eq.~(2.24) of Ref.~\cite{PhysRevD.16.1762}) for the
decay rate is given by
\begin{equation}
 \Gamma =\left[\frac{B}{2\pi\hbar}\right]^{\half}e^{-B/\hbar}
  \left|
   \frac{\det'\left[-\der_t^2+V''(\bar{x})\right]}
   {\det\left[-\der_t^2+\omega^2\right]}\right|^{-\half}
  \!\!\!\!
  \times \left[1+\mathcal{O}(\hbar)\right],
  \label{CC_formula}
\end{equation}
where $\bar{x}$ denotes the (zero-temperature) bounce solution and $B$
is the bounce action. We would like to
emphasize three points on this formula:
\begin{enumerate}
 \item They consider a large but finite time interval $T$, which is
       eventually sent to infinity.
 \item The determinants are supposed to be defined in the finite time
       interval $T$ with the Dirichlet boundary conditions at the
       boundaries $t=\pm T/2$.
 \item The determinant $\det'$ is the one with the zero eigenvalue
       omitted, and the zero-mode integration is performed as a
       collective coordinate integration, leading to the factor
       $(B/2\pi\hbar)^{\half} T$.
\end{enumerate}

% Dirichlet b.c. ?
In the saddle-point approximation, they write a general path $x(t)$
satisfying the boundary conditions $x(-T/2)=x_i$ and $x(T/2)=x_f$ as
\begin{equation}
 x(t)=\bar{x}(t) +\sum_{n} c_n x_n(t),
  \label{expand_0}
\end{equation}
where
\begin{align}
 \int_{-T/2}^{T/2} dt x_n(t) x_m(t) &=\delta_{nm},
 \label{cc_norm}\\
  x_n(\pm T/2)&=0, \label{cc_dirichlet}
\end{align}
and the path integral measure $[dx]$ is defined as
\begin{equation}
 [dx] =\prod_n (2\pi\hbar)^{-\half} dc_n.
  \label{measure}
\end{equation}
The complete set $\{x_n\}$ is chosen as the set of eigenfunctions of
the fluctuation operator,
\begin{equation}
 \left[-\frac{d^2}{dt^2} +V''(\bar{x})\right] x_n =\lambda_n x_n.
\end{equation}
Therefore, the determinant $\det\left[-\der_t^2+V''(\bar{x})\right] =
\prod_n\lambda_n$ is defined with the Dirichlet boundary conditions
Eq.~\eqref{cc_dirichlet}.

Because $\bar{x}(t)$ is a solution of the (Euclidean) equation of
motion, $d\bar{x}/dt$ is the eigenfunction with zero
eigenvalue. The normalization~\eqref{cc_norm} gives
\begin{equation}
 x_1(t) = B^{-\half} \frac{d\bar{x}}{dt}.
\end{equation}
The infinitesimal change of the corresponding coefficient $c_1$ may be
related to the infinitesimal change of the time of the classical
solution. Thus $dt = dc_1 B^{-\half}$ and the integration with the
measure $(2\pi\hbar)^{-\half}dc_1$ can be done with the measure
$(B/2\pi\hbar)^{\half} dt$, leading to the factor
$(B/2\pi\hbar)^{\half}T$

% not satisfactory
The calculation described above is however not satisfactory; First of
all, the Dirichlet boundary conditions are not compatible with the time
translational invariance on which the collective coordinate method is
based. The translational invariance is broken by the very boundary
conditions. Second, the relation $dt = dc_1 B^{-\half}$ holds only in
the vicinity of $t=0$. A symptom is explained in Appendix.~B of
Ref.~\cite{PhysRevD.95.085011} for example; the range of the thus
defined coefficient $c_1$ is bounded, incompatible to the integration
from $-T/2$ to $T/2$ (eventually, from $-\infty$ to $+\infty$).

% what one should have done
In order to have exact translational invariance with a finite $T$, one
should impose the periodic boundary conditions (in the sense of
Sec.~\ref{sec:intro}) on $x(t)$; $x(-T/2) = x(T/2), \ \dot{x}(-T/2) =
\dot{x}(T/2)$. 

% new expansion
Consider the time-translated classical solution $\bar{x}(t-t_0)$, and
the expansion~\cite{PhysRevD.95.085011}
\begin{equation}
 x(t) =\bar{x}(t-t_0) +\sum_n \bar{c}_n \bar{x}_n(t-t_0),
  \label{expand_t0}
\end{equation}
instead of the expansion Eq.~\eqref{expand_0}, where the eigenfunctions
$\{\bar{x}_n\}$ now satisfy the periodic boundary conditions.

% change of variable
The infinitesimal change of $x(t)$ with respect to the infinitesimal
change to $t_0$ may be related to the infinitesimal change of $\bar{c}_1$,
\begin{align}
 dx(t) &= \bar{x}(t-t_0-dt_0)-\bar{x}(t-t_0) =
 dt_0 \frac{d\bar{x}}{dt}(t-t_0) 
 \notag \\
  &=d\bar{c}_1 \bar{x}_1(t-t_0) = d\bar{c}_1 B^{-\half}
 \frac{d\bar{x}}{dt}(t-t_0),
\end{align}
we have $dt_0 = d\bar{c}_1 B^{-\half}$. This relation holds for any
value of $t_0 \in [-T/2,T/2]$. Thus the integration over $\bar{c}_1$ may
be replaced by the integration over $t_0$. That is, instead of
Eq.~\eqref{measure}, we have
\begin{equation}
 [dx] = (B/2\pi\hbar)^{\half} dt_0 
  \prod_{n\ne1} (2\pi\hbar)^{-\half} d\bar{c}_n.
  \label{cc_collective}
\end{equation}
The integration over $t_0$ leads to the factor $(B/2\pi\hbar)^{\half}T$.

% what CC was supposed to calculate
It is therefore not the amplitude
\begin{equation}
 \langle 0|e^{-HT/\hbar}|0\rangle
  =\mathcal{N} \int_{x(\pm T/2)=0} [dx]\; e^{-S[x]/\hbar},
\end{equation}
which Callan and Coleman pretended to calculate (with the Dirichlet
boundary conditions $x_i=x_f=0$), but a
path-integral with the periodic boundary conditions
\begin{equation}
 \mathcal{N} \int_{\text{p.b.c.}}[dx]\; e^{-S[x]/\hbar},
  \label{pbc}
\end{equation}
that one should be calculating with the method of collective
coordinates. However, the operator representation corresponding to
Eq.~\eqref{pbc} is obscure to us.

% it is not a trace
Note that Eq.~\eqref{pbc} is not the trace
\begin{align}
 \tr(e^{-HT/\hbar}) &=\int dy\; \langle y|e^{-HT/\hbar}|y\rangle
 \notag \\
 &=\int dy\; \mathcal{N} \int_{x(\pm T/2)=y} [dx]\; e^{-S[x]/\hbar}.
 \label{trace}
\end{align}
Eq.~\eqref{pbc} does not include a shot-type path $\dot{x}(\pm
T/2)\ne0, \ \dot{x}(T/2)=-\dot{x}(-T/2)$, while Eq.~\eqref{trace}
does.

% determinant should be defined with PBC
The determinant of the fluctuation operator should also be defined with
the periodic boundary conditions. It contains zero eigenvalue even
for a finite $T$. 

% the final expression is unchanged, but the BC changed
According to the approach described above, the final result that Callan
and Coleman should have obtained is given by the same expression
Eq.~\eqref{CC_formula} but now the determinants are defined with the
periodic boundary conditions, and the $T\to \infty$ limit is understood.

% still not completely satisfactory
In this subsection, we pointed out the problem related to the
translational invariance and the collective coordinate, and suggested a
viable approach to solve it. Even if we take the approach, however, it
is not completely satisfactory. The bounce solution at zero temperature
does not exactly vanish for a finite $T$, so that it has somewhere a
point where $\dot{\bar{x}}(t)$ is not smooth and thus $\bar{x}(t)$ is
not a classical solution there. To have classical solutions smooth
everywhere, one should consider classical solutions with the period $T$,
which is in essence the finite-$T$ bounce and the shifted bounce
solutions considered in Sec~\ref{sec:bounce}. The best way of obtaining
the decay rate at zero temperature is, in our opinion, to obtain the
finite-temperature result and take the zero-temperature limit of
it. That is our next task.

\subsection{Zero-temperature limit of $\ZB$}
\label{subsec:zeroTbounce}

% zero-temperature limit
In the zero-temperature limit ($\beta \to \infty$), $\xb\to \xF = 0$,
$\xr \to \xR$, and $E \to 0$. Since $\SB$ has a smooth limit (which we
call $\SB^{0}$), we only need to know how $d\beta/dE$ in
Eq.~\eqref{ZB_final} behaves in the zero-temperature limit.

% d\beta/dxb from the period 
The period $\beta$ may be written as
\begin{equation}
  \beta=2 \int_{\xb}^{\xr} \frac{dx}{\sqrt{2(V(x)-V(\xb))}},
   \label{beta}
\end{equation}
since $\dot{x}=\pm\sqrt{2(V(x)+E)}=\pm\sqrt{2(V(x)-V(\xb))}$. When
$\beta$ is very large, $\xb$ is very close to $\xF$, so that for $x$
close to $\xb$, the potential $V(x)$ may be well approximated by a
parabola,
\begin{equation}
 V(x) \approx V(\xb)+V'(\xb)(x-\xb)+\half V''(\xb)(x-\xb)^2.
\end{equation}
Having this in mind, we rewrite Eq.~\eqref{beta} as
\begin{align}
 \beta &=2\int_{\xb}^{\xr}dx
 \bigg[
 \frac{1}{\sqrt{2(V(x)-V(\xb))}}
 \notag \\
 &\qquad\qquad\quad
 -\frac{1}{\sqrt{2V'(\xb)(x-\xb)+V''(\xb)(x-\xb)^2}}
 \bigg]
 \notag \\
 &+2\int_{\xb}^{\xr}dx
 \frac{dx}{\sqrt{2V'(\xb)(x-\xb)+V''(\xb)(x-\xb)^2}}.
\end{align}
The first integral now has a smooth limit when $\xb \to \xF$. The second
integral is elementary;
\begin{align}
& \int_{\xb}^{\xr}dx
 \frac{dx}{\sqrt{2V'(\xb)(x-\xb)+V''(\xb)(x-\xb)^2}}
\notag \\
 =& \ 
 \frac{1}{\sqrt{V''(\xb)}}\ln
 \frac{X+\sqrt{X^2-(V'(\xb)/V''(\xb))^2}}{(V'(\xb)/V''(\xb))},
\end{align}
where $X =\xr-\xb+V'(\xb)/V''(\xb)$.

% the priod in the limit
For large $\beta$, we get
\begin{align}
 \beta &\approx \frac{2}{\sqrt{V''(\xF)}}\int_{\xF}^{\xR}dx
  \left[
   \frac{\sqrt{V''(\xF)}}{\sqrt{2(V(x)-V(\xF))}}-\frac{1}{x-\xF}
  \right]
 \notag \\
 &+\frac{2}{\sqrt{V''(\xF)}}\ln\left[\frac{2(\xR-\xF)}{\xb-\xF}\right],
 \label{largebetaB}
\end{align}
thus
\begin{align}
 &\lim_{\beta\to\infty}(\xb-\xF)^2e^{\omega\beta}
 \notag \\
 =&\;4(\xR-\xF)^2e^{2\int_{\xF}^{\xR}dx
  \left[
   \frac{\omega}{\sqrt{2(V(x)-V(\xF))}}-\frac{1}{x-\xF}
  \right]
  },
 \label{limbeta}
\end{align}
where we introduce $\omega = \sqrt{V''(\xF)}$. Note that the right-hand
side of Eq.~\eqref{limbeta} is a constant. 
%
%{\color{red} 
Noting $\ln(\xb-\xF) =
\ln\left((\xb-\xF)e^{\omega\beta/2}\right)-\omega \beta/2 $ and using
l'H\^opital's rule, we obtain
\begin{equation}
 \lim_{\beta\to \infty}\frac{d\beta/d\xb}{(\xb-\xF)^{-1}} 
  =\lim_{\beta\to\infty}\frac{\beta}{\ln(\xb-\xF)} 
  =-\frac{2}{\omega},
\end{equation}
where we used $\lim_{\beta\to\infty}\ln((\xb-\xF)e^{\omega\beta/2})$ is
finite due to Eq.~\eqref{limbeta}.
%
%} 
%By differentiating it with respect to $\xb$, 
We thus find
\begin{equation}
 \lim_{\beta\to\infty} (\xb-\xF)\frac{d\beta}{d\xb}=-\frac{2}{\omega}.
  \label{limxb}
\end{equation}

% dbeta/dE in terms of dbeta/dxb
We also note
\begin{equation}
 \frac{d\beta}{dE} =\frac{d\beta}{d\xb}\frac{d\xb}{dE}
  =-\frac{d\beta}{d\xb}\frac{1}{V'(\xb)},
\end{equation}
where we have used $V'(\xb)(d\xb/dE)=-1$, which follows from
$V(\xb)=-E$.  When $\beta$ is very large, $\xb$ is very close to $\xF$,
thus $V(\xb)\approx V(\xF)+ \half \omega^2(\xb-\xF)^2$. It follows
$V'(\xb)\approx \omega^2(\xb-\xF)$.

% zero-temperature limit of ZB
Therefore, we find
\begin{equation}
 \lim_{\beta\to \infty} \frac{d\beta}{dE} 
  = \lim_{\xb\to\xF}\frac{2\omega}{\omega^4(\xb-\xF)^2}.
  \label{dbetadE}
\end{equation}
By substituting it into Eq.~\eqref{ZB_final}, we finally get
\begin{equation}
 \lim_{\beta\to\infty} \ZB/\beta
  = \lim_{\xb\to\xF} \frac{i}{\sqrt{2\pi}} e^{-\SB^{0}}
  \frac{\omega^2(\xb-\xF)}{\sqrt{2\omega}}.
  \label{lim_ZB}
\end{equation}

\subsection{Zero-temperature limit of $\ZsF$}
\label{subsec:zeroTshot}

% zero temperature limit
In the zero-temperature limit ($\beta\to\infty$), $\xt$ and $\xb$ both
go to $\xR$, so that the imaginary branch solution $\xsF^{(\text{I})}$
does not exist, and we only have the real branch contribution.  In this
subsection, we omit the superscript (R) for notational simplicity. The
turning point $\xs$ goes to $\xF$. The action has a smooth limit, which
we call $\SsF^0(y)$.  Therefore we only need to know how
$\sqrt{\left|d\SsF/dE\right|} = \sqrt{2\mu_{\text{s}}(-\beta/2;y)
\nu_{\text{s}}(-\beta/2;y)}$ behaves to get the limit.

% nu_s may be obtained from dy/dxs
Let us start with $\nu_{\text{s}}(-\beta/2;y)
=\der_{E}\xsF(-\beta/2;y)$. Since $y=\xsF(-\beta/2;y)$, one may write it
as
\begin{equation}
 \nu_{\text{s}}(-\beta/2;y) =\frac{d\xs}{dE}\frac{dy}{d\xs} 
  =-\frac{1}{V'(\xs)}\frac{dy}{d\xs},
  \label{nu_s}
\end{equation}
since $-V(\xs)=E$.

% dy/dxs from the period
The period $\beta$ is given by
\begin{equation}
 \beta =2\lim_{\epsilon\to 0}\int^y_{\xs+\epsilon} 
  \frac{dx}{\sqrt{2(V(x)-V(\xs))}},
  \label{betaSF}
\end{equation}
where a small positive number $\epsilon$ is introduced. By keeping the
period fixed and differentiating it with respect to $\xs$, we have
\begin{align}
 0&= 2\lim_{\epsilon\to0}\bigg[
 \frac{dy}{d\xs} \frac{1}{\sqrt{2(V(y)-V(\xs))}}
 \notag \\
 &\qquad\qquad
 -\frac{1}{\sqrt{2(V(\xs+\epsilon)-V(\xs))}}
 \bigg]
 \notag \\
 &\qquad\qquad
 +\int^y_{\xs+\epsilon}
 \frac{V'(\xs)dx}{\left[2(V(x)-V(\xs))\right]^{\frac{3}{2}}}.
\end{align}
By adding and substituting $1/\sqrt{2(V(y)-V(\xs))}$, we rewrite it as
\begin{align}
 &\left(\frac{dy}{d\xs}-1\right)\frac{1}{\sqrt{2(V(y)-V(\xs))}}
 \notag \\
  =&\lim_{\epsilon\to 0}\int^y_{\xs+\epsilon}
 \frac{V'(x)-V'(\xs)}{\left[2(V(x)-V(\xs))\right]^{\frac{3}{2}}}dx,
\end{align}
thus, 
\begin{align}
 \lim_{\beta\to\infty}\frac{dy}{d\xs} 
 &=\half\lim_{\xs\to\xF}\int^y_{\xs}
  \frac{\sqrt{V(y)-V(\xF)} \; V'(x)}{(V(x)-V(\xF))^{\frac{3}{2}}}dx+1
 \notag \\
 &=\lim_{\xs\to\xF}\sqrt{\frac{V(y)-V(\xF)}{V(\xs)-V(\xF)}}.
\end{align}
By substituting it into Eq.~\eqref{nu_s}, we obtain
\begin{equation}
 \lim_{\beta\to\infty} \nu_s(-\beta/2)
  =-\lim_{\xs\to\xF}\frac{1}{V'(\xs)}
  \sqrt{\frac{V(y)-V(\xF)}{V(\xs)-V(\xF)}}
  \label{nu_s_lim}
\end{equation}

% mu_s
Let us move to $\mu_{\text{s}}(-\beta/2;y) =
\dot{x}_{\text{sF}}(-\beta/2;y)$. From the energy conservation, we
have
\begin{equation}
 \mu_{\text{s}}(-\beta/2;y)
  =-\sqrt{2(V(y)-V(\xs))}.
  \label{mu_s_lim}
\end{equation}
Note the sign of the right-hand side.

% the zero-tempearature limit of ZsF
By combining Eq.~\eqref{nu_s_lim} and Eq.~\eqref{mu_s_lim}, we obtain
\begin{align}
 &\lim_{\beta\to\infty}
   \left|\frac{d\SsF(y)}{dE}\right|
 \notag \\
 =& \lim_{\beta\to\infty}
 \frac{2\sqrt{2(V(y)\!-\!V(\xs))(V(y)\!-\!V(\xF))}}
 {V'(\xs)\sqrt{V(\xs)-V(\xF)}}
 \notag \\
 =& \lim_{\beta\to \infty}
 \frac{4(V(y)-V(\xF))}{\omega^3(\xs-\xF)^2},
\end{align}
and it leads to
\begin{equation}
 \lim_{\beta\to\infty} \ZsF = \lim_{\xs\to\xF}
  \half
  \sqrt{\frac{1}{2\pi\omega}} \int_{\xF}^{\xR}dy 
  \frac{\omega^2(\xs-\xF) e^{-\SsF^0(y)}}{\sqrt{V(y)-V(\xF)}}.
  \label{lim_ZsF}
\end{equation}

\subsection{Zero-temperature limit of the decay rate}
\label{zeroTGamma}

% (xb-xF)/(xs-xF)
In order to compute the zero-temperature limit of the decay rate, we
need to know the relation between $(\xb-\xF)$ in Eq.~\eqref{lim_ZB} and
$(\xs-\xF)$ in Eq.~\eqref{lim_ZsF}. For this purpose, let us return to
Eq.~\eqref{betaSF}.  In a parallel way to getting Eq.~\eqref{limbeta},
we have
\begin{align}
 &\lim_{\beta\to\infty}(\xs-\xF)^2e^{\omega\beta}
 \notag \\
  =&\ 
   4(y-\xF)^2
   e^{2\int_{\xF}^{y}dx
   \left[\frac{\omega}{\sqrt{2(V(x)-V(\xF))}}-\frac{1}{x-\xF}\right]}.
 \label{limbeta2}
\end{align}
Taking the ratio of Eq.~\eqref{limbeta} to Eq.~\eqref{limbeta2},
we obtain
\begin{equation}
 \lim_{\beta\to\infty}\frac{\xb-\xF}{\xs-\xF} =
  e^{\int_{y}^{\xR}
  \frac{\omega dx}{\sqrt{2(V(x)-V(\xF))}}}.
  \label{lim_diffs}
\end{equation}

% the final result
Now, we are ready to calculate the zero-temperature limit of the
false-vacuum decay rate,
\begin{align}
 \lim_{\beta\to\infty} \Gamma &\approx
  \lim_{\beta\to\infty}\frac{1}{\beta}\frac{\text{Im} \ZB}{\ZsF}
 \notag \\
 &\approx 
 \frac{e^{-\SB^{0}}}
 {\int_{\xF}^{\xR} dy
 \frac{e^{-\SsF^0(y)}}
 {\sqrt{2(V(y)-V(\xF))}}
 e^{-\int_{y}^{\xR}
 \frac{\omega dx}{\sqrt{2(V(x)-V(\xF))}}}}.
 \label{zeroTlimit}
\end{align}

% CC formula rewritten
This is different from the CC formula~\eqref{CC_formula}. Before
comparing our result with it, let us rewrite it in a convenient form. As
we show in Appendix~\ref{app:det}, the ratio of the determinants in the
$T\to\infty$ limit is the same both for the periodic boundary conditions
and the Dirichlet boundary conditions, and is given by
Eq.~\eqref{zerolimpbc}.  Thus, ignoring $\mathcal{O}(\hbar)$ and using
our notation, Eq.~\eqref{CC_formula} may be written as
\begin{equation}
 \Gamma_{\text{CC}} =\sqrt{\frac{\omega}{\pi}}\omega
  (\xR\!-\!\xF) 
  e^{-\SB^0}
  e^{\int_{\xF}^{\xR}dx
  \left[\frac{\omega}{\sqrt{2(V(x)-V(\xF))}}-\frac{1}{x-\xF}\right]}.
  \label{decayCC}
\end{equation}

% it is related to our formula
Interestingly, these two are not completely unrelated. Consider the
integrand of the denominator of Eq.~\eqref{zeroTlimit},
\begin{equation}
 X(y) \equiv \frac{e^{-\SsF^0(y)}}
 {\sqrt{2(V(y)-V(\xF))}}
 e^{-\int_{y}^{\xR}dx\left[\frac{\omega
 dx}{\sqrt{2(V(x)-V(\xF))}}\right]}.
\end{equation}
In the limit $y\to\xF$, $\SsF^0(y) \to 0$,
$\sqrt{2(V(y)-V(\xF))} \to \omega(y-\xF)$, we may obtain
\begin{equation}
 X^{-1}(\xF) =\omega(\xR-\xF) 
  e^{\int_{\xF}^{\xR}dx\left[\frac{\omega
 dx}{\sqrt{2(V(x)-V(\xF))}}-\frac{1}{x-\xF}\right]}.
\end{equation}
Therefore, we find that $\Gamma_{\text{CC}}$ may be written in terms of
$X(\xF)$ as
\begin{equation}
 \Gamma_{\text{CC}} =\frac{e^{-\SB^0}}{\sqrt{\frac{\pi}{\omega}}X(\xF)}.
\end{equation}
The factor $\sqrt{\pi/\omega}$ corresponds to the $y$-integration. For a
typical potential, $X(y)$ has a peak near $y=\xF$, but the peak is not
sharp. In the next section, we compare our decay rate with
$\Gamma_{\text{CC}}$ as well as the WKB decay rate $\Gamma_{\text{WKB}}$
obtained in Ref.~\cite{PhysRevD.95.085011}.

\section{numerical solutions and the comparison with previous works}
\label{sec:num}

In this section, we perform some numerical calculations for several
potentials. 

\subsection{Decay rates as a function of the potential in the
  zero-temperature}

We first compare our zero-temperature limit~\eqref{zeroTlimit} with
$\Gamma_{\text{CC}}$~\eqref{decayCC} and NLO WKB decay rate
$\Gamma_{\text{WKB}}$ (Eq.(2.19) in Ref.~\cite{PhysRevD.95.085011}),
which may be written as
\begin{equation}
 \Gamma_{\text{WKB}} =
  \frac{e^{-2\int_a^b dx \pi(x)}}
  {\int_{\xF}^a \frac{dx}{p(x)}
  +\int_a^b \frac{dy}{\pi(y)}e^{-2\int_a^ydx \pi(x)}},
\end{equation}
where $p(x)=\sqrt{2(E_0-V(x))}$, $\pi(x)=\sqrt{2(V(x)-E_0)}$, and $E_0$
is the ``ground-state'' energy, $\omega/2$. The classical turning
points $a$ and $b$ ($\xF < a < b < \xR$) satisfy $V(a)=V(b)=E_0$.

\begin{figure}[h]
 \includegraphics[width=\linewidth,clip]{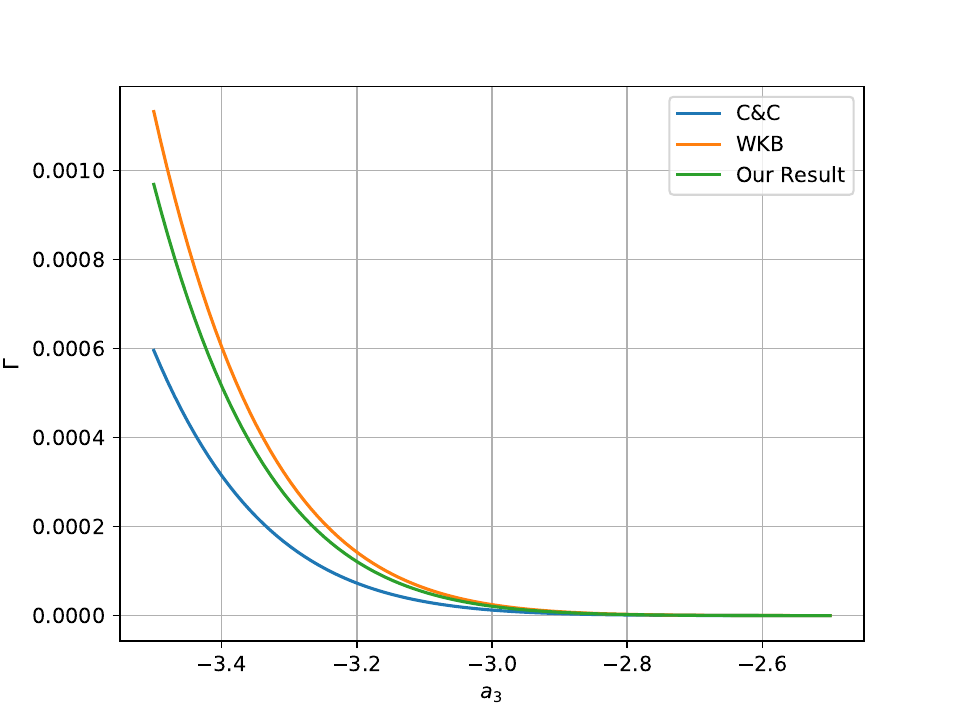}
 \caption{\label{fig:decay_rate_x3} Decay rates for the potential
 $V_1(x)= a_3x^3+8x^2$. The zero-temperature limit of our
 result~\eqref{zeroTlimit}, $\Gamma_{\text{CC}}$, and
 $\Gamma_{\text{WKB}}$ are plotted as functions of the parameter $a_3$,
 which controls the potential barrier.}
\end{figure}

We consider the potential
\begin{equation}
 V_1(x) = a_3 x^3+8x^2\quad x\in [\xR,\xF],
\end{equation}
which satisfies $V(0)=V'(0)=0$, $V''(0)=16$ ($\omega=4$) and the
parameter $a_3$ controls the potential barrier. Note that the three
formulae depend only on the shape of the potential for $x\in[\xF,
\xR]$. It is necessary for the height of the potential $V_{\text{max}}$
is much larger than the energy $\omega/2$, that would be the ground
state energy if the tunneling were turned off. The range of $a_3$ must
be chosen to satisfy this constraint.  In Fig.~\ref{fig:decay_rate_x3},
we compare the three decay rates. The decay rates decrease as the
potential height increases, as expected. Our result is similar to the
WKB result in both magnitude and slope, but quite different from the CC
result.

\begin{figure}[h]
 \includegraphics[width=\linewidth,clip]{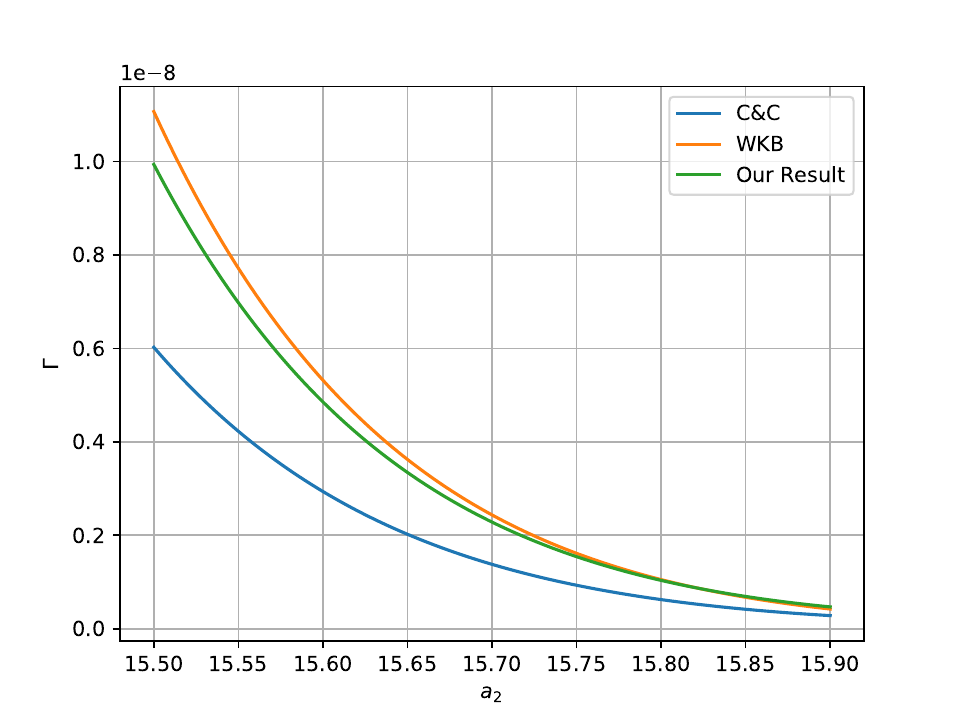}
 \caption{\label{fig:decay_rate_almost_sym} Decay rates for the
 potential $V_2(x)= x^4-8x^3+a_2x^2$. The zero-temperature limit of our
 result~\eqref{zeroTlimit}, $\Gamma_{\text{CC}}$, and
 $\Gamma_{\text{WKB}}$ are plotted as functions of the parameter $a_2$,
 when the potential is close to the symmetric case.}
\end{figure}

Fig.~\ref{fig:decay_rate_almost_sym} shows the decay rates for another
potential
\begin{equation}
 V_2(x) = x^4-8x^3+a_2x^2,
\end{equation}
where the parameter $a_2$ is close to $16$, that is, the potential is
close to the symmetric case $V_{\text{sym}}(x)=x^2(x-4)^2$. Again, the
zero-temperature limit of our result is close to the WKB result, but
different from the CC result.

\begin{figure}[h]
 \includegraphics[width=\linewidth,clip]{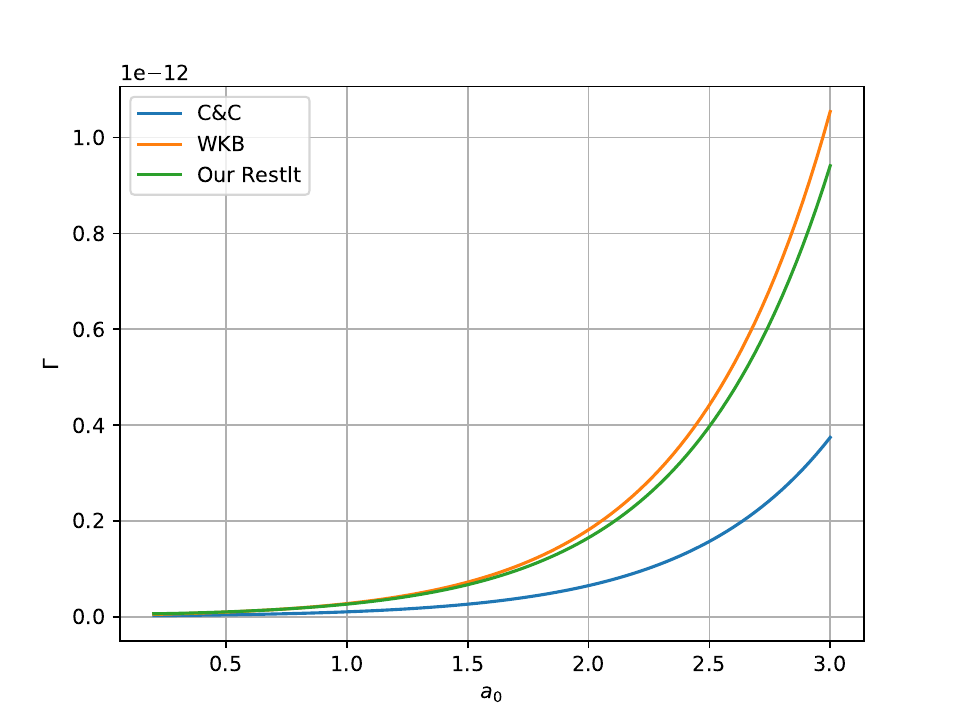}
 \caption{\label{fig:cosh_pot_at_zeroT} Decay rates for the potential
 $V_{\text{np}}(x)= \cosh(0.2 x)+10/\cosh(x+a_0)$. The
 zero-temperature limit of our result~\eqref{zeroTlimit},
 $\Gamma_{\text{CC}}$, and $\Gamma_{\text{WKB}}$ are plotted as
 functions of the parameter $a_0$.}
\end{figure}

%
%{\color{red}
In Fig.~\ref{fig:cosh_pot_at_zeroT}, we show the decay rates for even
another potential of non-polynomial type,
\begin{equation}
 V_{\text{np}} =\cosh(0.2 x) +\frac{10}{\cosh(x+a_0)}.
\end{equation}
The result is very similar to that of
Fig.~\ref{fig:decay_rate_almost_sym}, for a wide range of the parameter
$a_0$.
%
%}

\subsection{Temperature dependence of the decay rate}
\label{subsec:T-dep}

In this subsection, we numerically compute our result
Eq.~\eqref{our_result} as a function of $\beta^{-1}$ and compare it
with that of Affleck's formula, obtained from Eq.(14) in
Ref.~\cite{PhysRevLett.46.388} with $\Gamma =(2/\hbar)\; \text{Im}
F$. It is given in our notation by
\begin{equation}
 \Gamma_{\text{Affleck}}
  =\frac{\left|\left|\dotxB\right|\right|}{\sqrt{2\pi}}
  \left|
   \frac{\det\left[-\frac{d^2}{d\tau^2}+\omega^2\right]}
   {\det'\left[-\frac{d^2}{d\tau^2}+V''(\xB(\tau))\right]}
  \right|^{\half}_{\text{p.b.c.}} e^{-\SB},
  \label{affleck}
\end{equation}
where the determinants are defined in the finite interval $\beta$.
Note that the zero-temperature limit of $\Gamma_{\text{Affleck}}$ is
$\Gamma_{\text{CC}}$.

The ratio of the determinants defined with the periodic boundary
conditions is given by Eq.~\eqref{ratio_finiteT_periodic}.  By
substituting it, Eq.~\eqref{affleck} may be written as
\begin{equation}
 \Gamma_{\text{Affleck}} =
  \frac{1}{\sqrt{2\pi}}2\sinh(\omega\beta/2)e^{-\SB}
  \left[\frac{d\beta}{dE}\right]^{-\half}.
  \label{affleck2}
\end{equation}

We first describe the method of the numerical calculations briefly.

In order to obtain $\ZB$, we start with $\xb$ (and $\xr$ obtained by
$V(\xb) = V(\xr)$) and solve Eq.~\eqref{beta} to determine the period
$\beta$. In this way, we have sets of $(\xb,\beta)$. The energy $E$ is
also determined by $\xb$ through $E=-V(\xb)$. A small change $\Delta
\xb$ leads to small changes $\Delta \beta$ and $\Delta E$, so that we
can calculate $\Delta \beta/\Delta E$ numerically. Of course one may
use a more elaborate method to calculate $d\beta/dE$.

The calculation of $\ZsF^{(\text{R})}$ and $\ZsF^{(\text{I})}$ is more
complex. Starting with $\xs$, one needs to find $y$ that satisfies
Eq.~\eqref{betaSF} for a given $\beta$. The energy $E$ is determined
by $\xs$ through $E=-V(\xs)$. As explained in
Sec.~\ref{sec:shot}, there are two solutions for $y \in
(\xr,\xt)$. These two solutions have different values of the action,
$\SsF^{(\text{R})}(y)$ and $\SsF^{(\text{I})}(y)$.

Numerical calculations get more difficult as we go closer to the zero
temperature limit, because $d\beta/dE$ diverges exponentially.

\begin{figure}[h]
 \includegraphics[width=\linewidth,clip]{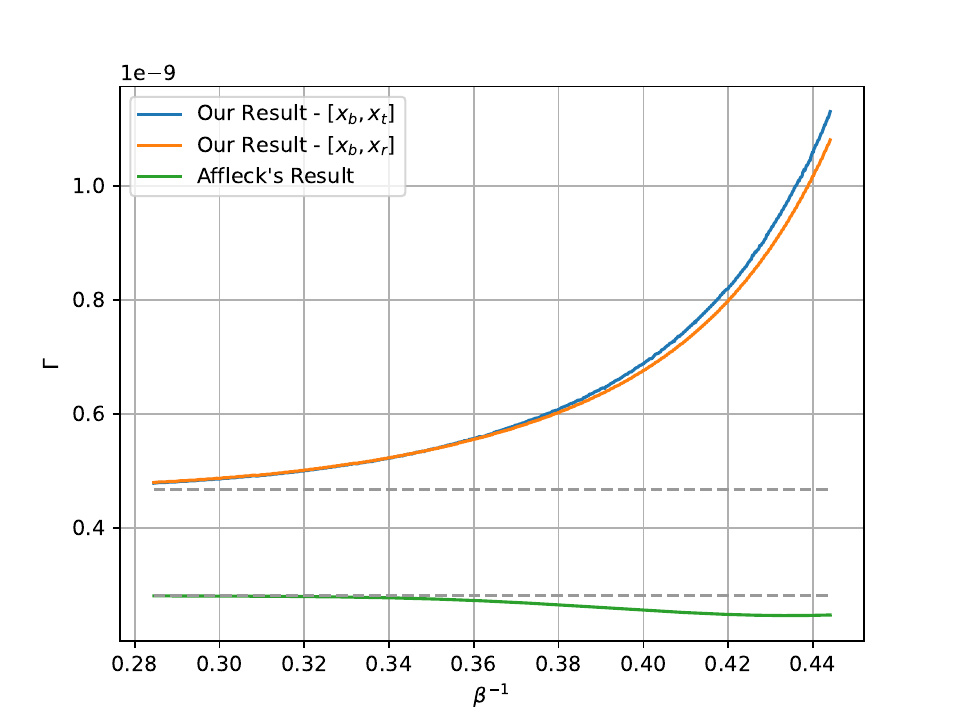}
 \caption{\label{fig:decay_rate_finite_T} Decay rates for the potential
 $V_3(x)= x^4-8x^3+15.9x^2$. Our result~\eqref{our_result} with full
 contributions from $y\in [\xb, \xt]$, our result
 only with the contributions from $y\in [\xb, \xr]$, and Affleck's
 result~\eqref{affleck} are plotted as functions of $\beta^{-1}$. The
 dashed lines denote the zero-temperature limits.}
\end{figure}

\begin{figure}[h]
 \includegraphics[width=\linewidth,clip]{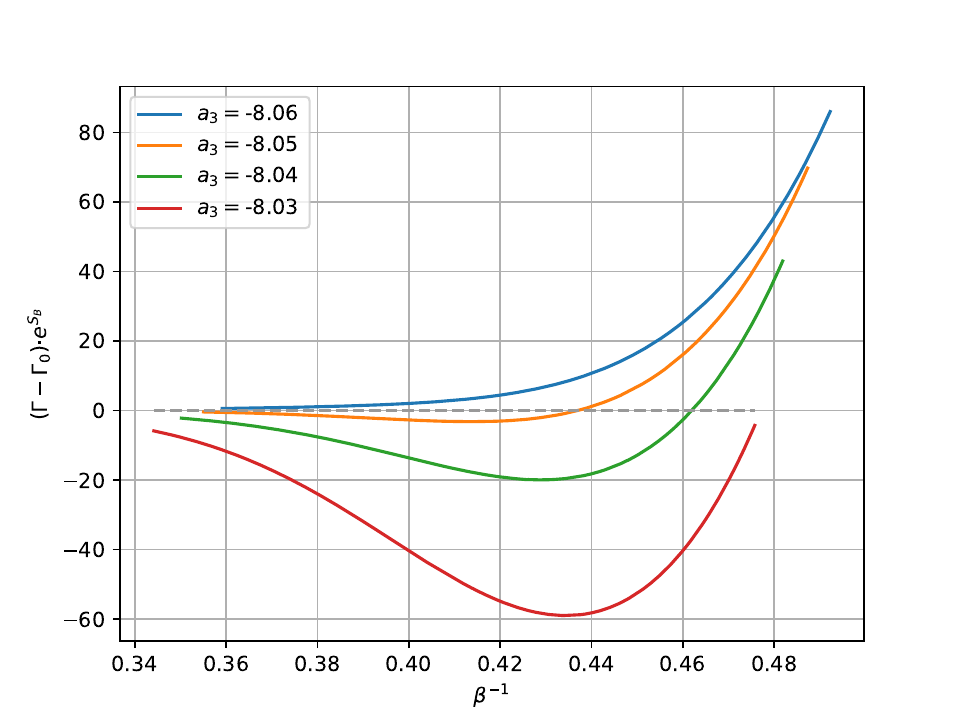}
 \caption{\label{fig:decay_rate_Affleck} Affleck's decay rates for the
 potential $V_4(x)= x^4+a_3x^3+16x^2$ is plotted as a function of
 $\beta^{-1}$. $\Gamma_0$ is the zero-temperature limit of
 $\Gamma_{\text{Affleck}}$. The parameter of the potential is close to
 the symmetric case $a_3=-8$. }
\end{figure}

Fig.~\ref{fig:decay_rate_finite_T} shows how the decay rates change with
temperature for the potential
\begin{equation}
 V_3(x) = x^4-8x^3+15.9x^2,
\end{equation}
which is close to the symmetric potential. Our result shows a monotonous
increase while Affleck's result shows a dip before it starts to
increase. We think that there is no physical reason for the dip. It
occurs when the potential is very close to the symmetric double well
potential. Fig.~\ref{fig:decay_rate_Affleck} shows Affleck's result for
the potential
\begin{equation}
 V_4(x) = x^4+a_3x^3+16x^2.
\end{equation}
We see that as $V_4(x)\to V_{\text{sym}}(x)\ (a_3\to-8)$, the dip gets deeper.

We understand why this happens.  When the potential is symmetric, the
system spends a lot of time not only near $\xb$ but also near
$\xr$. Thus $d\beta/dE$ is much larger than that for a generic
case. Actually, for a symmetric double well potential $V(x)$, the period
$\beta$ may be written as
\begin{equation}
 \beta = 4 \int_{\xb}^{\xC} \frac{dx}{\sqrt{2(V(x)-V(\xb))}},
\end{equation}
where $\xC =(\xF+\xR)/2$. Thus, instead of
Eq.~\eqref{limxb}, we have
\begin{equation}
 \lim_{\beta\to\infty} (\xb-\xF)\frac{d\beta}{d\xb}=-\frac{4}{\omega},
  \label{dbetadxb_sym}
\end{equation}
and it leads to
\begin{align}
 &\lim_{\beta\to\infty}\frac{d\beta}{dE} 
 \notag \\
 =& \ \lim_{\beta\to\infty}
 \frac{e^{\omega\beta/2}}{\omega^3(\xC-\xF)^2}
  e^{-2\int_{\xF}^{\xC} dx
  \left[\frac{\omega}{\sqrt{2(V(x)-V(\xF))}}-\frac{1}{x-\xF}\right]},
 \label{limbeta_sym}
\end{align}
thus for large $\beta$,
\begin{equation}
 \Gamma_{\text{Affleck}}\sim \sinh(\omega\beta/2)
  \left[\frac{d\beta}{dE}\right]^{-\half}
  \sim e^{\omega\beta/4}.
\end{equation}
It diverges in the zero-temperature limit. Since the zero-temperature
limit of $\Gamma_{\text{Affleck}}$ is $\Gamma_{\text{CC}}$, the decay
rate of Callan and Coleman is also divergent for a symmetric potential.

When the potential is close to (but not) symmetric,
$\Gamma_{\text{Affleck}}$ is finite but large in the zero-temperature
limit. As the temperature increases from zero, $\Gamma_{\text{Affleck}}$
first decreases since $\Gamma_{\text{Affleck}}\sim e^{\omega\beta/4}$ is
approximately correct.

Our result also shows an steep increase when $\beta^{-1} = 0$, but it
does not diverge. Let us look at the denominator of
Eq.~\eqref{zeroTlimit},
\begin{equation}
 \text{Denom.} =\int_{\xF}^{\xR} \!\!dy
 \frac{e^{-\SsF^0(y)}}
 {\sqrt{2(V(y)\!-\!V(\xF))}}
 e^{-\int_{y}^{\xR}\!\!
 \frac{\omega dx}{\sqrt{2(V(x)\!-\!V(\xF))}}}.
\end{equation}
By integrating by parts, we have
\begin{align}
\text{Denom.} &=\frac{1}{\omega}
 \left[
 e^{-\SsF^0(y)}
 e^{-\int_{y}^{\xR}\!\!\frac{\omega dx}{\sqrt{2(V(x)\!-\!V(\xF))}}}
 \right]^{\xR}_{\xF}
 \notag \\
 &\ +\frac{1}{\omega}\int_{\xF}^{\xR} \!\!dy
 \frac{d\SsF^{0}(y)}{dy}
 e^{-\SsF^0(y)}
 e^{-\int_{y}^{\xR}\!\!\frac{\omega dx}{\sqrt{2(V(x)\!-\!V(\xF))}}}
 \notag \\
 &=\frac{1}{\omega}
 e^{-\SB^{0}}
 \notag \\
  &\ +\frac{1}{\omega}\int_{\xF}^{\xR} \!\!dy
 \frac{d\SsF^{0}(y)}{dy}
 e^{-\SsF^0(y)}
 e^{-\int_{y}^{\xR}\frac{\omega dx}{\sqrt{2(V(x)\!-\!V(\xF))}}},
 \label{denom}
\end{align}
where we have used $\SsF^{0}(\xR)=\SB^{0}$ and $e^{-\omega\beta}\to 0$.
Because of the singularity at $\xR$ the integral in the exponential
diverges except for $y=\xR$, and
\begin{equation}
 \left.\frac{d\SsF^{0}(y)}{dy}\right|_{y=\xR} =0,
\end{equation}
we see the second term of Eq.~\eqref{denom} vanishes. Thus, for a
symmetric double well potential, we have
\begin{equation}
 \lim_{\beta\to\infty} \Gamma = \omega.
\end{equation}
This is of course an unphysical result.

\begin{figure}[h]
 \includegraphics[width=\linewidth,clip]{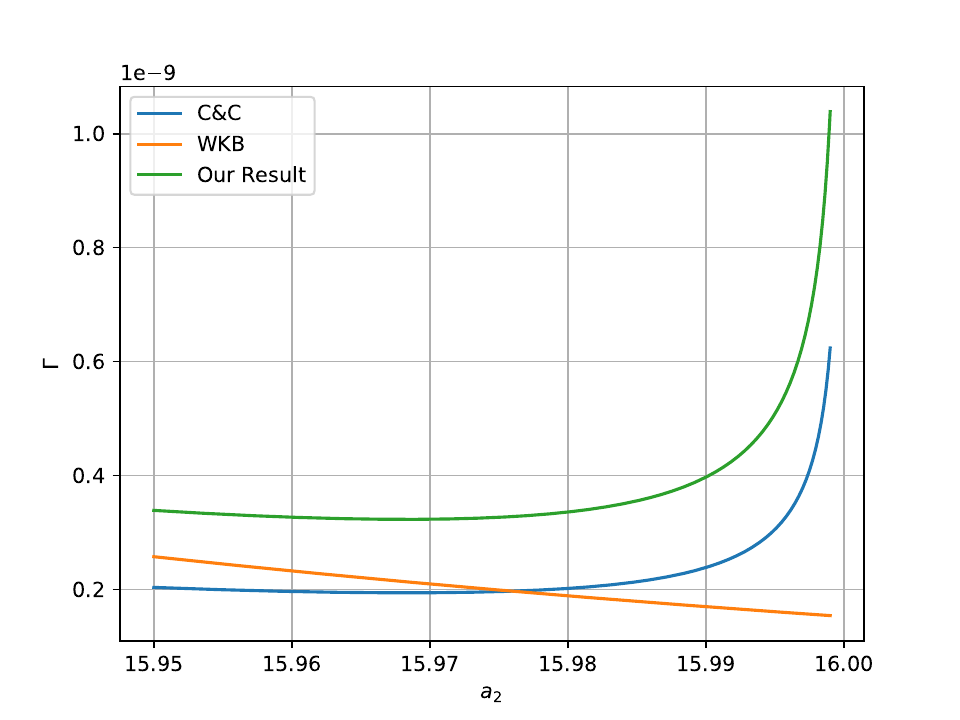}
 \caption{\label{fig:decay_rate_very_close} Decay rates for the
 potential $V_2(x) =x^4-8x^3+a_2x^2$. The zero-temperature limit of our
 result~\eqref{zeroTlimit}, $\Gamma_{\text{CC}}$, and
 $\Gamma_{\text{WKB}}$ are plotted as functions of the parameter $a_2$,
 when the potential is very close to the symmetric limi ($a_2\to
 16$). See Fig.~\ref{fig:decay_rate_almost_sym}.}
\end{figure}

At zero-temperature, our decay rate and $\Gamma_{\text{CC}} =
\lim_{\beta\to \infty}\Gamma_{\text{Affleck}}$ both show a steep
increase when the potential becomes very close to the symmetric one, as
shown in Fig.~\ref{fig:decay_rate_very_close}. There, our method somehow
breaks down and fails to give a reasonable answer.

%
%{\color{red} 
The failure of the calculations seems to be due to the splitting of a
bounce into a pair of an instanton and an anti-instanton when
$V'(\xR)=0$. Namely, the particle travels from $x=\xF$ to $x=\xR$ and
spend an arbitrary long time there and comes back to the original
position. Thus the negative eigenvalue decreases in magnitude and
becomes a quasi-zero mode (with the maginitude of the eigenvalue being
exponentially small), associated with the distance between the
instanton and the anti-instanton. Note that the (original) zero mode is
now associated with the center of the pair. On the other hand, Callan
and Coleman (as well as the present authors) did not take into account
the quasi-zero mode.
%
%}

As we showed in Fig.~\ref{fig:decay_rate_finite_T} and
Fig.~\ref{fig:decay_rate_Affleck}, it affects the finite-temperature
behavior of $\Gamma_{\text{Affleck}}$ even when the potential is not
very close to the symmetric one. However, it does not seem to affect
that of our decay rate. The difference probably comes from the fact that
$\Gamma_{\text{Affleck}}$ diverges as $\beta\to \infty$, while the
limit of our decay rate stays finite.

\section{summary}
\label{sec:summary}

In this paper, we considered a one-dimensional quantum-mechanical system
in a potential which has a metastable false-vacuum at finite
temperature, and calculated the decay rate in the saddle-point
approximation.

Because of the trace, the static solutions do not contribute and the
shot-F solution play an important role. The trace integral combines
all the contributions from the shifted bounce solutions.

There are two shot-F solutions for $y\in(\xr,\xt)$, one of which is
imaginary. It plays a similar role to the finite-$T$ bounce and
shifted bounce solutions in the calculation of the decay rate.

We then calculated the zero-temperature limit of our result. We also
commented on the finite time ``regularization'' and the collective
coordinate method of Callan and Coleman.

Finally, we performed some numerical calculations. We first showed the
potential dependence of our decay rate and compared it with that of
Callan and Coleman, and with that obtained by the NLO WKB
approximation. We find that our result is similar to the WKB one. Next,
we showed the temperature dependence of our decay rate and compared it
with that of Affleck. We showed that Affleck's decay rate first
decreases as temperature increases and then increases when the potential
is close to a symmetric double well potential. This behavior seems to
have no physical reasons, and seems to be a consequence of the
divergence of the decay rate for a symmetric double well potential at
zero-temperature limit. The zero-temperature limit of our decay rate
also shows a steep increase when the potential is very close to a
symmetric double well, it does not seem to affect the temperature
dependence when the potential is not symmetric. 

It is interesting to extend our results to quantum field theory in which
the contributions from shot-type solutions have not been
included. Investigation in this direction is now under consideration.

\appendix
\section{Dimensionless variables}
\label{app:dim}

In order to simplify the expressions and to do numerical calculations,
we use dimensionless variables. In this Appendix, we summarize how
they are defined.

The Hamiltonian of a one-dimensional quantum mechanics of a particle of
mass $m$ is given by
\begin{equation}
 H=\frac{p^2}{2m} +V(x),
\end{equation}
where the momentum $p$ and the coordinate $x$ have physical dimensions.
The inverse temperature $\beta=1/k_{\text{B}}T$ has the inverse
dimension of energy.

We introduce an energy scale $\epsilon_0$ and by using it, we define
dimensionless variables.

We define a dimensionless coordinate $\mathcal{X} =
\frac{\sqrt{m\epsilon_0}}{\hbar}x$ and a dimensionless momentum
$\mathcal{P} = \frac{1}{\sqrt{m\epsilon_0}}p$, as well as a
dimensionless (Euclidean) time $\mathcal{T} =
\frac{\epsilon_0}{\hbar}\tau$. They satisfy the following relation,
\begin{equation}
 \frac{d\mathcal{X}}{d\mathcal{T}} = \mathcal{P}.
\end{equation}
The Hamiltonian may be written as
\begin{align}
 H&=\epsilon_0 \mathcal{H}(\mathcal{P}, \mathcal{X}), \\
 \mathcal{H}&=\half \mathcal{P}^2+\mathcal{V}(\mathcal{X}),
\end{align}
where $\mathcal{H}$ is the dimensionless Hamiltonian, and we have
defined a dimensionless potential
\begin{equation}
 \mathcal{V}(\mathcal{X})
  =\epsilon_0^{-1}
  V\left(
    \frac{\hbar}{\sqrt{m\epsilon_0}}\mathcal{X}
   \right).
\end{equation}
We can also define a dimensionless inverse temperature
\begin{equation}
 \mathcal{B}=\beta\epsilon_0.
\end{equation}
A dimensionless free energy defined as
$\mathcal{F}=\epsilon_0^{-1}F$, may be expressed as
\begin{equation}
 \mathcal{F} =-\frac{1}{\mathcal{B}}\ln
  \tr\left(e^{-\mathcal{B}\mathcal{H}}\right).
\end{equation}
The physical decay rate $\Gamma$ may be obtained from the imaginary
part of a certain combination of contributions to $\mathcal{F}$, by
multiplying a factor $\epsilon_0/\hbar$.

\section{Ratio of the determinants defined with periodic boundary
 conditions}
\label{app:det}

In this Appendix, we explain how to calculate the ratio of the
determinants defined with the periodic boundary conditions, which is
useful to understand Eq.~\eqref{CC_formula} and Eq.~\eqref{affleck}.

The basic formula is given by Forman~\cite{Forman_1897, Forman_1992},
which is a generalization of Gel'fand-Yaglom theorem. See also
Ref.~\cite{Dunne_lecture}.

We consider two differential operators
\begin{equation}
 \mathcal{O}^{(i)}=-\frac{d^2}{d\tau^2}+W^{(i)}(\tau)\quad (i=1,2)
\end{equation}
in the interval $\tau \in [-\beta/2, \beta/2]$, and the eigenvalue
problems;
\begin{equation}
 \mathcal{O}^{(i)} \phi_n^{(i)}(\tau) = \lambda_n^{(i)} \phi_n^{(i)}(\tau).
\end{equation}
Now the eigenfunctions are supposed to satisfy the conditions
\begin{equation}
 M
  \begin{pmatrix}
   \phi_n^{(i)}(-\beta/2) \\
   \dot{\phi}_n^{(i)}(-\beta/2)
  \end{pmatrix}
  +N
  \begin{pmatrix}
   \phi_n^{(i)}(\beta/2) \\
   \dot{\phi}_n^{(i)}(\beta/2)
  \end{pmatrix}
  =
  \begin{pmatrix}
   0 \\
   0
  \end{pmatrix}
  \quad (i=1,2), 
\end{equation}
where $M$ and $N$ are $2\times2$ matrix. For the periodic boundary
conditions, 
\begin{equation}
 M=
  \begin{pmatrix}
   1 & 0 \\
   0 & 1
  \end{pmatrix},
  \qquad
  N=
  \begin{pmatrix}
   -1 & 0 \\
   0 & -1 
  \end{pmatrix}.
  \label{pbc_MN}
\end{equation}
The determinant $\det\mathcal{O}^{(i)}$ is (naively) an infinite
product of the eigenvalues,
\begin{equation}
 \det\mathcal{O}^{(i)}=\prod_{n} \lambda_n^{(i)}.
\end{equation}

As we did in Sec.~\ref{sec:bounce} for Gel'fand-Yaglom theorem, we
consider the zero-eigenvalue problem for each operator,
\begin{equation}
 \mathcal{O}^{(i)} \psi^{(i)}_j(\tau) =0 \quad (j=1,2),
\end{equation}
where $\psi^{(i)}_j$ are two independent solutions, and consider the
$2\times2$ matrices
\begin{equation}
 H^{(i)}(\tau)=\begin{pmatrix}
    \psi^{(i)}_1(\tau) & \psi^{(i)}_2(\tau) \\
    \dot{\psi}^{(i)}_1(\tau) & \dot{\psi}^{(i)}_2(\tau)
   \end{pmatrix}, 
\end{equation}
and
\begin{equation}
 Y^{(i)}(\tau) =H^{(i)}(\tau){H^{(i)}}^{-1}(-\beta/2)
\end{equation}
Forman's theorem says that
\begin{equation}
 \frac{\det\mathcal{O}^{(1)}}{\det\mathcal{O}^{(2)}}
  = \frac{\det_{2\times2}(M+NY^{(1)}(\beta/2))}
  {\det_{2\times2}(M+NY^{(2)}(\beta/2))},
  \label{forman}
\end{equation}
where on the right-hand side, $\det_{2\times2}$ denotes the determinant for a
$2\times2$ matrix.

If we choose $\psi^{(i)}_j(\tau)$ so that $H^{(i)}(-\beta/2)=
\begin{pmatrix}
 1 & 0 \\
 0 & 1
\end{pmatrix}$, then we have $Y^{(i)}(\tau)=H^{(i)}(\tau)$ and the
 expression is simplified considerably. We assume that this choice of
 $\psi^{(i)}_j(\tau)$ has been done.

Note that for the Dirichlet boundary conditions,
\begin{equation}
 M=\begin{pmatrix}
    1 & 0 \\
    0 & 0
   \end{pmatrix}, \qquad
   N=\begin{pmatrix}
      0 & 0 \\
      1 & 0
     \end{pmatrix},
\end{equation}
so that
\begin{equation}
 \frac{\det \mathcal{O}^{(1)}}
  {\det \mathcal{O}^{(2)}}=
  \frac{\psi^{(1)}_2(\beta/2)}
  {\psi^{(2)}_2(\beta/2)},
\end{equation}
with the boundary conditions $\psi_2^{(i)}(-\beta/2)=0$ and
$\dot{\psi}_2^{(i)}(-\beta/2)$. This is nothing but Gel'fand-Yaglom
theorem for the Dirichlet boundary conditions, considered in
Sec.~\ref{sec:bounce}.

For the periodic boundary conditions~\eqref{pbc_MN},
Eq.~\eqref{forman} can be written as
\begin{equation}
  \left.
   \frac{\det\mathcal{O}^{(1)}}{\det\mathcal{O}^{(2)}}
  \right|_{\text{p.b.c.}}
  = \left.
     \frac{(1-\psi^{(1)}_1)(1-\dot{\psi}^{(1)}_2)-\dot{\psi}^{(1)}_1\psi^{(1)}_2}
  {(1-\psi^{(2)}_1)(1-\dot{\psi}^{(2)}_2)-\dot{\psi}^{(2)}_1\psi^{(2)}_2}
  \right|_{\tau=\frac{\beta}{2}}.
  \label{ratio_periodic}
\end{equation}

Let us now consider the specific example
$\mathcal{O}^{(1)} = -\frac{d^2}{d\tau^2}+\omega^2$ and
$\mathcal{O}^{(2)} = -\frac{d^2}{d\tau^2}+V''(\xB(\tau))$. One can
easily get
\begin{align}
 \psi^{(1)}_1(\tau) &=\cosh\left(\omega(\tau+\beta/2)\right), \\
 \psi^{(1)}_2(\tau) &=\omega^{-1}\sinh\left(\omega(\tau+\beta/2)\right).
\end{align}
From these, we immediately obtain
\begin{equation}
 \left.
  (1-\psi^{(1)}_1)(1-\dot{\psi}^{(1)}_2)-\dot{\psi}^{(1)}_1\psi^{(1)}_2
  \right|_{\tau=\frac{\beta}{2}}
 = -4\sinh^2(\omega\beta/2).
\end{equation}
For the operator $\mathcal{O}^{(2)}$, $\psi^{(2)}_j(\tau)$ may be
written as linear combinations of $\mu(\tau)$ and $\nu(\tau)$ as in
Sec.~\ref{sec:bounce} (but defined with $\tau_c=0$). We get
\begin{align}
 \psi^{(2)}_1(\tau)
 &=\dot{\nu}(-\beta/2)\mu(\tau)-\dot{\mu}(-\beta/2)\nu(\tau), \\
 \psi^{(2)}_2(\tau)
 &= -\nu(-\beta/2)\mu(\tau)+\mu(-\beta/2)\nu(\tau).
\end{align}

With the periodic boundary conditions, $\det\mathcal{O}^{(2)}$
contains a zero eigenvalue to be removed. The way of removing the zero
mode is similar to the case with Dirichlet boundary conditions
considered in Sec.~\ref{sec:bounce}. We define $\det'\mathcal{O}^{(2)}
=\lim_{\lambda\to0} \det[\mathcal{O}^{(2)}-\lambda]/(-\lambda)$, and
consider $\psi^{(2)}_{\lambda j}\ (j=1,2)$ which satisfies
\begin{equation}
 \left[\mathcal{O}^{(2)}-\lambda\right]\psi^{(2)}_{\lambda j}(\tau)=0,
\end{equation}
and the initial value conditions $H_{\lambda}^{(2)}(-\beta/2)=
\begin{pmatrix}
 1 & 0 \\
 0 & 1
\end{pmatrix}$. As we did in Eq.~\eqref{to_get_lambda}, we consider 
\begin{align}
 0&= \int_{-\beta/2}^{\beta/2} d\tau \; \dotxB(\tau)
 \left[\mathcal{O}^{(2)}-\lambda\right]\psi^{(2)}_{\lambda j}(\tau)
 \notag \\
 &= \left[-\dotxB(\tau)\dot{\psi}^{(2)}_{\lambda j}
 +\ddotxB(\tau)\psi^{(2)}_{\lambda j}(\tau)\right]_{-\beta/2}^{\beta/2}
 \notag \\
 &\quad 
 -\lambda\int_{-\beta/2}^{\beta/2} d\tau\;
 \dotxB(\tau)\psi^{(2)}_{\lambda j}(\tau).
\end{align}
By substituting the initial values and using the periodicity, we obtain
\begin{align}
 &
 \begin{pmatrix}
  \psi^{(2)}_{\lambda 1}(\beta/2)-1 & -\dot{\psi}^{(2)}_{\lambda
  1}(\beta/2) \\
  \psi^{(2)}_{\lambda 2}(\beta/2) & 1-\dot{\psi}^{(2)}_{\lambda 2}(\beta/2)
 \end{pmatrix}
 \begin{pmatrix}
  \ddotxB(\beta/2) \\
  \dotxB(\beta/2)
 \end{pmatrix}
 \notag \\
=& \lambda
 \begin{pmatrix}
  \int_{-\beta/2}^{\beta/2} d\tau\; \dotxB(\tau)\psi^{(2)}_{\lambda
  1}(\tau) \\
  \int_{-\beta/2}^{\beta/2} d\tau\; \dotxB(\tau)\psi^{(2)}_{\lambda
  2}(\tau) 
 \end{pmatrix},
\end{align}
and thus
\begin{align}
 &\begin{pmatrix}
  \ddotxB(\beta/2) \\
  \dotxB(\beta/2)
 \end{pmatrix}
 = \frac{-\lambda}{-D}
 \begin{pmatrix}
  1-\dot{\psi}^{(2)}_{\lambda 2}(\beta/2) & \dot{\psi}^{(2)}_{\lambda
  1}(\beta/2) \\
  -\psi^{(2)}_{\lambda 2}(\beta/2) & \psi^{(2)}_{\lambda 1}(\beta/2)-1 
 \end{pmatrix}
 \notag \\
 &\times 
 \begin{pmatrix}
  \int_{-\beta/2}^{\beta/2} d\tau\; \dotxB(\tau)\psi^{(2)}_{\lambda
  1}(\tau) \\
  \int_{-\beta/2}^{\beta/2} d\tau\; \dotxB(\tau)\psi^{(2)}_{\lambda
  2}(\tau) 
 \end{pmatrix},
\end{align}
where
\begin{align}
 -D&= \left(1-\psi^{(2)}_{\lambda 1}(\beta/2)\right)
  \left(1-\dot{\psi}^{(2)}_{\lambda 2}(\beta/2)\right)
 \notag \\
 &\qquad 
 -\dot{\psi}^{(2)}_{\lambda 1}(\beta/2)\psi^{(2)}_{\lambda 2}(\beta/2)
\end{align}
is the combination that appears in
Eq.~\eqref{ratio_periodic}. 

Therefore, we obtain
\begin{align}
 &
 \left.
 \frac{\det \mathcal{O}^{(1)}}{\det'\mathcal{O}^{(2)}}
 \right|_{\text{p.b.c.}}
 = \lim_{\lambda\to0}
 \left.
 \frac{\det \mathcal{O}^{(1)}}
 {\det\left[\mathcal{O}^{(2)}-\lambda \right]/(-\lambda)}
 \right|_{\text{p.b.c.}} 
 \notag \\
=&\ 
 \left[-4\sinh^2(\omega\beta/2)\right]\lim_{\lambda\to0}\frac{-\lambda}{-D}.
\end{align}
The limit may be written as
\begin{align}
& \frac{\ddotxB(\frac{\beta}{2})}
  {\int_{-\frac{\beta}{2}}^{\frac{\beta}{2}}\!d\tau \dotxB(\tau)\!
  \left[
   (1\!-\!\dot{\psi}^{(2)}_{2}(\frac{\beta}{2}))\psi^{(2)}_1(\tau)
   \!+\!\dot{\psi}^{(2)}_1(\frac{\beta}{2})\psi^{(2)}_2(\tau)
  \right]}\notag \\
=& -\frac{\ddotxB(\beta/2)}
 {\left[\ddotxB(\beta/2)\right]^2\left|\left|\dotxB\right|\right|^2
 \nu(\beta/2)}
 \left[\frac{d\beta}{dE}\right]^{-1}
 \notag \\
=& \frac{1}{\left|\left|\dotxB\right|\right|^2}
 \left[\frac{d\beta}{dE}\right]^{-1},
\end{align}
where we have used
\begin{align}
 \psi^{(2)}_{1}(\beta/2)
 &=\left[\ddotxB(\beta/2)\right]^2\frac{d\beta}{dE}, \\
 \psi^{(2)}_{2}(\beta/2)
 &= 1,
\end{align}
and $\ddotxB(\beta/2)\nu(\beta/2)=-1$. We finally obtain
\begin{equation}
 \left.
 \frac{\det \mathcal{O}^{(1)}}{\det'\mathcal{O}^{(2)}}
 \right|_{\text{p.b.c.}}
 =-\frac{4\sinh^2(\omega\beta/2)}
 {\left|\left|\dotxB\right|\right|^2}
 \left[\frac{d\beta}{dE}\right]^{-1}.
 \label{ratio_finiteT_periodic}
\end{equation}

To conclude this appendix, we will show that the zero-temperature limit
of the ratio of the determinants with the periodic boundary conditions
is equal to that with the Dirichlet boundary conditions,
\begin{equation}
 \lim_{\beta\to\infty}
  \left.
   \frac{\det \mathcal{O}^{(1)}}{\det'\mathcal{O}^{(2)}}
  \right|_{\text{p.b.c.}}
  =\lim_{\beta\to\infty}
  \left.
   \frac{\det \mathcal{O}^{(1)}}{\det'\mathcal{O}^{(2)}}
  \right|_{\text{d.b.c}}.
\end{equation}
In the zero-temperature limit, $\left|\left|\dotxB\right|\right|^2\to
\SB^{0}$, and from Eq.~\eqref{dbetadE} and Eq.~\eqref{limbeta}, we see 
\begin{align}
 &\lim_{\beta\to \infty}
  \left.
   \frac{\det \mathcal{O}^{(1)}}{\det'\mathcal{O}^{(2)}}
  \right|_{\text{p.b.c.}} \notag \\
 =& -\lim_{\beta\to\infty} 
 \frac{e^{\omega \beta}}{\SB^{0}}\frac{\omega^4(\xb-\xF)^2}{2\omega}
 \notag \\
 =&-\frac{2\omega^3}{\SB^{0}}(\xR-\xF)^2
 e^{2\int_{\xF}^{\xR}dx
  \left[
   \frac{\omega}{\sqrt{2(V(x)-V(\xF))}}-\frac{1}{x-\xF}
  \right]
  }.
 \label{zerolimpbc}
\end{align}
On the other hand, from Eq.~\eqref{ratio_of_dets}, by using
$\left|\ddotxB(\beta/2)\right|^2 =
\left|V'(\xb)\right|^2=\omega^4(\xb-\xF)^2$, we have
\begin{align}
 &\lim_{\beta\to\infty}
  \left.
   \frac{\det \mathcal{O}^{(1)}}{\det'\mathcal{O}^{(2)}}
  \right|_{\text{d.b.c}}
 \notag \\
 =& -\lim_{\beta\to\infty} \frac{\omega^4(\xb-\xF)^2}{\SB^{0}} 
 \frac{e^{\omega\beta}}{2\omega}
 \notag \\
 =& -\frac{2\omega^3}{\SB^{0}}(\xR-\xF)^2
 e^{2\int_{\xF}^{\xR}dx
  \left[
   \frac{\omega}{\sqrt{2(V(x)-V(\xF))}}-\frac{1}{x-\xF}
  \right]
  }.
\end{align}
This completes the proof.

% If you have acknowledgments, this puts in the proper section head.
\begin{acknowledgments}
 The work of Q.Y. was supported by Kyushu University Leading Human
 Resources Development Fellowship Program (Quantum Science Area).
\end{acknowledgments}

% Create the reference section using BibTeX:
\bibliography{./FalseVacuum}

\end{document}